\documentclass[12pt]{JHEP3} 
\usepackage{epsfig}
\usepackage{amsmath,amssymb}

\newcommand{\startappendix}{
\setcounter{section}{0}
\renewcommand{\thesection}{\Alph{section}}}

\newcommand{\Appendix}[1]{
\refstepcounter{section}
\begin{flushleft}
{\large\bf Appendix \thesection: #1}
\end{flushleft}}

\newcommand{\be}{\begin{equation}}
\newcommand{\ee}{\end{equation}}
\newcommand{\bear}{\begin{eqnarray}}
\newcommand{\eear}{\end{eqnarray}}
\newcommand{\ba}{\begin{array}}
\newcommand{\ea}{\end{array}}

\newcommand{\CN}{{\cal N}} 
 \newcommand{\CW}{{\cal W}}
\newcommand{\CG}{{\cal G}}
\newcommand{\Tr}{{\rm Tr}}

\newcommand{\HCG}{{\hat{\cal G}}}

\newcommand{\balpha}{{\boldsymbol{\alpha}}}
\newcommand{\bzeta}{{\boldsymbol{\zeta}}}
\newcommand{\bbeta}{{\boldsymbol{\beta}}}
\newcommand{\bsigma}{{\boldsymbol{\sigma}}}

\newcommand{\bvarphi}{{\boldsymbol{\varphi}}}
\newcommand{\brho}{{\boldsymbol{\rho}}}
\newcommand{\bxi}{{\boldsymbol{\xi}}}

\newcommand{\bH}{{\boldsymbol{H}}}
\newcommand{\bw}{{\boldsymbol{w}}}
\newcommand{\bz}{{\boldsymbol{z}}}
\newcommand{\bX}{{\boldsymbol{X}}}
\newcommand{\bbe}{{\boldsymbol{e}}}
\newcommand{\bv}{{\boldsymbol{v}}}
\newcommand{\CI}{{\cal I}}

\def\tr{{\rm tr}}

\preprint{KIAS-P04014
\\SNUTP04-001 \\ UTIG-02-04\\ hep-th/0403076}

\title{The $\CN=1^*$ Theories on $R^{1+2}\times S^1$ with Twisted
Boundary Conditions}

\author{Seok Kim$^1$, Ki-Myeong Lee$^{2,3}$, Ho-Ung Yee$^2$, and Piljin
Yi$^2$
\\
\\
$^1$School of Physics, Seoul National University, Seoul 151-747,
Korea
\\$^2$Korea Institute for Advanced Study, Seoul 130-722, Korea\\
$^{3\dagger}$Physics Department, University of Texas at Austin,
Texas 78712,
USA\\
Email: calaf2@snu.ac.kr, klee@kias.re.kr, ho-ung.yee@kias.re.kr,
piljin@kias.re.kr \\
$^\dagger${\small\rm (until the end of May, 2004)}}

\abstract{We explore the $\CN=1^*$ theories compactified on a
circle with twisted boundary conditions. The gauge algebra of
these theories are the so-called twisted affine Lie algebra. We
propose the exact superpotentials by guessing the sum of all
monopole-instanton contributions and also by requiring $SL(2,Z)$
modular properties. The latter is inherited from the $\CN=4$
theory, which will be justified in the M theory setting.
Interestingly all twisted theories possess full $SL(2,Z)$
invariance, even though none of them are simply-laced. We further
notice  that these superpotentials are associated with certain
integrable models widely known as elliptic Calogero-Moser models.
Finally, we argue that the glueball superpotential must be
independent of the compactification radius, and thus of the
twisting, and confirm this by expanding it in terms of glueball
superfield in weak coupling expansion.}

\begin{document}
\baselineskip 18pt

\section{Introduction}

The $\CN=1^*$ theory is obtained by adding masses for all three
adjoint chiral supermultiplets of the $\CN=4$ supersymmetric
theory. It is well known that the classical vacuum structure of
the $\CN=1^*$ theory is characterized by embedding of $SU(2)$
algebra to the gauge algebra \cite{Vafa:1994tf}. Quantum
mechanically, however, the phenomena of gaugino condensation and
chiral symmetry breaking further complicate the physics of vacua.
The Seiberg-Witten curve for the $\CN=2^*$ theory, where only two
out of three chiral multiplets of the $\CN=4$ theory have nonzero
equal mass,  was found early on in the study of Seiberg-Witten
theory \cite{Seiberg:1994aj,Donagi:1995cf}. Interestingly, the
singularities of these curves were found to determine the vacuum
structure of the corresponding $\CN=1^*$ theory. Furthermore, it
has been shown that the quantum $\CN=1^*$ theory with $SU(N)$
gauge group inherits the $SL(2,Z)$ symmetry of the $\CN=4$ theory,
whereby the massive vacua of the quantum theory are transformed to
each other\cite{Donagi:1995cf}.

On the other hand, the Seiberg-Witten curves of pure $\CN=2$ super
Yang-Mills theories with any semi-simple Lie group have been found
to have a deep connection to classical integrable models. The
Seiberg-Witten curves for the $\CN=2$ pure Yang-Mills theory turn
out to be  the spectral curves of the Toda-models
\cite{Donagi:1995cf,Gorsky:1995zq,Martinec:1995by,
Martinec:1995qn,Marshakov:1996nv,Nakatsu:1995bz}\footnote{ The
connection between $\CN=2$ theories and integrable models can be
also realized in string theory
\cite{Witten:1997sc,Katz:1997eq,Gorsky:1996fu}.}. Furthermore, the
Seiberg-Witten curves for $\CN=2^*$ theories with semi-simple Lie
algebra are related to the so-called twisted elliptic
Calogero-Moser models
\cite{Martinec:1995qn,Itoyama:1995nv,D'Hoker:1997ha,D'Hoker:1998yi}.

The study of supersymmetric theories  compactified on a circle was
initiated in Ref.~\cite{Seiberg:1996nz}. For the $\CN=1$ pure
Yang-Mills theory which is periodic on the circle,  the  gauge
fields in the Cartan part of the gauge algebra may be dualized to
chiral fields, and the supersymmetric effective potential as a
function of these chiral fields has been obtained and found to be
the twisted affine Toda
potential~\cite{katzvafa,Vafa:1998vs,Davies:1999uw,Davies:2000nw}.
For the case of the $\CN=1^*$ $SU(N)$  theory compactified on a
circle with $SU(N)$ gauge group, the effective superpotential was
argued to be the potential of the elliptic $SU(N)$ Calogero-Moser
model,  on which  the $SL(2,Z)$ acts as a modular
transformation~\cite{Dorey:1999sj}. For $\CN=1^*$ theories with
other simple gauge groups with periodic boundary condition, there
is some work on the superpotential \cite{Kumar:2001iu}, but it has
not been fully explored.

In this paper, we wish to study the structure of $\CN=1^*$
theories with arbitrary semi-simple gauge group compactified on a
circle, but with a twist. When one of  spatial directions is
compactified on a circle, one may impose  twisted boundary
condition instead of periodic one, without breaking supersymmetry.
Twisted boundary condition here means \footnote{Stringy
realization of such a twist has been considered in
Ref.~\cite{Bershadsky:1996nh,Park:1996it}.}
\begin{equation}
\Phi(x_4+2\pi R)=\bsigma[\Phi(x_4)]\,,
\end{equation}
where $\bsigma$ is an outer automorphism acting on the adjoint
representation with the property $\bsigma^L=1$ for some integer
$L$, and $x_4$ is the compactified direction of radius $R$.
Throughout this paper, we will adopt the notation, $\CG^{(L)}$,
for the theory obtained by twisting  the corresponding $\CN=1^*$
theory based on Lie algebra $\CG$. The twistable $\CG$'s are
exhausted by  $A_r$, $D_r$, and  $E_6$. We wish to find exact
superpotentials of these theories by extending  the above
intricate relationships found on the theories with periodic
boundary condition. A preliminary consideration of this problem
has appeared in Ref.~\cite{Hanany:2001iy}.

A much simpler version of this problem would be to consider the
limit of the pure $\CN=1$ theories on $R^{1+2}\times S^1$. This is
obtained from $\CN=1^*$ by letting all chiral multiplets
infinitely massive. With periodic boundary condition and arbitrary
simple gauge group, the effective superpotential has been
calculated and found to be exactly the potential of the Toda model
associated with the $\CN=2$ theory
\cite{katzvafa,Vafa:1998vs,Davies:1999uw,Davies:2000nw}. This Toda
potential has a nice physical interpretation as arising from a
supersymmetric generalization of the Polyakov mechanism, namely as
nonperturbative contributions from fundamental monopole instantons
on Euclidean three dimensional space time.

One nice aspect of this simpler $\CN=1$ theory compactified on a
circle is that the superpotential truncates down to a sum over
``unit" monopole-instanton contributions. The usual semiclassical
computation with the dilute gas approximation suffices. When one
introduces twisted boundary condition along the circle, the pure
$\CN=1$ theory can be treated easily by using the same approach as
in Ref.~\cite{Davies:1999uw,Davies:2000nw}. In weak coupling
limit, one calculate the effective  superpotential from the
contributions from fundamental monopole instantons.  Our result on
the effective superpotential for the $\CN=1$ theories for all
semi-simple gauge groups with twisted boundary condition also are
given by the potential of  the Toda-type model with definite
normalization. The Toda model is associated to the twisted
affine-algebra.

The superpotential turns out to be  holomorphic and independent of
the compactification radius and so  is valid in any radius or
coupling. However, the boundary condition becomes irrelevant in
the infinite radius limit. Thus, the physics of ground states,
vacuum degeneracy and the gluino condensation, should be identical
to that of the theories with  periodic boundary condition. This
turns out to be borne out with help of the three rather nontrivial
identities satisfied by the roots of the untwisted and twisted
Kac-Moody algebra.

With three massive chiral multiplets in the adjoint representation
in the $\CN=1^*$ theory, however, this nice Toda-like feature is
no longer exact. One must sum over all possible combinations of
monopole instantons to obtain the exact answer. Approaching this
problem in the usual dilute gas approximation is obviously
ill-fated, and we must turn to other methods. We first guess that
the summation of all monopole instantons of a given type can be
written in terms of the Weierstrass elliptic function  as in the
untwisted $SU(N)$ case of Ref.~\cite{Dorey:1999sj}. We will use
the consistency with  small coupling limit to the Toda potential
and   Weyl symmetry to reach a preliminary form of the
superpotential.

The most powerful condition  that will lead us to the correct
answer is the $SL(2,Z)$ symmetry. As was mentioned already, the
$SL(2,Z)$ of $\CN=4$ $SU(N)$ theories is known to be inherited by
their $\CN=1^*$ cousins. We will assume that this inheritance is
generic, and proceed by determining  the $SL(2,Z)$ modular
properties of the $\CN=4$ theories with  twisted boundary
condition. Then the problem is reduced to picking out appropriate
superpotential which is  modular under $SL(2,Z)$ transformations.

The middle part of this paper will discuss in much detail the
issue of the $SL(2,Z)$ in $\CN=4$ theories with twisted boundary
condition, where we will make a crucial use of  M-theoretical and
stringy realizations of the twisted $\CN=4$ theories. At the end
of this discussion, we will find that all ``exact''
superpotentials correspond to the potentials of elliptic
Calogero-Moser models, once again demonstrating the close
connection between Seiberg-Witten theories, compactified $\CN=1$
theories, and integrable models.

Recall that the integrable models related to $\CN=2^*$ theories
have been found to be the elliptic Calogero-Moser models
\cite{Martinec:1995qn,D'Hoker:1997ha}, or rather, to be more
precise, the corresponding integrable models are determined by the
co-root vectors of the gauge group, leading to the twisted
Calogero-Moser models \cite{D'Hoker:1998yi}. Thus, given the usual
relationship between $\CN=2$ Seiberg-Witten curve and $\CN=1$
superpotential, the effective superpotentials for $\CN=1^*$
theories with  periodic boundary condition are expected to be
related to the integrable models of twisted elliptic
Calogero-Moser models \cite{D'Hoker:1998yi}.

In contrast, for the theory with twisted boundary condition, we
find that the effective superpotentials are related to the
untwisted elliptic Calogero-Moser models, $C_r$, $B_r$, $F_4$, and
$G_2$ types, respectively, for the group $A_{2r-1}^{(2)}$,
$D_{r+1}^{(2)}$, $E_6^{(2)}$, and $D_4^{(3)}$, and a twisted
elliptic Calogero-Moser model of $BC_r$ type for the group
$A_{2r}^{(2)}$. These integrable models have been studied in
Ref.\cite{D'Hoker:1998yi,Inozemtsev,Bordner:1998sw}.

Our computation, which covers both twisted and untwisted cases, is
actually more specific than those in the literature, as we have
unambiguous normalization factors for every term in the
superpotential, including one which only depends on the coupling
constant. We fixed them utilizing small coupling limit, Weyl
symmetry, and modular properties under $SL(2,Z)$. This allows us
the detailed exploration of the full modular property of the
effective potential. We would like to take this as an independent
evidence for the validity of our approach.

We subject our effective potential to another nontrivial
consistency check. The effective superpotential turns out to be
independent of the compactification radius. While this does not
immediately imply that the physics is independent of boundary
condition, we can deduce that the physics of gaugino condensation
and vacua labelled by this order parameter are independent of the
boundary condition. This independence may be rather surprising
since the shape of superpotential does depend on whether the
theory is twisted or not.

For this, we introduce the glueball superfield, $S\sim \Tr
W^\alpha W_\alpha$, {\it a la} Veneziano and Yankielowicz
\cite{Veneziano:1982ah}, and extract the glueball superpotential.
This is achieved by a `dual' or `mirror' transformation in weak
coupling series, which is a kind of Legendre
transformation~\cite{Hori:2000kt}. The leading term of the
superpotential, which is the Toda potential of the $\CN=1$ theory,
is mapped to the Veneziano-Yankielowicz potential for the glueball
superfield after this dual transformation. Higher order monopole
instanton corrections of the $\CN=1^*$ theory are mapped to higher
order terms in the glueball superpotential.\footnote{The $\CN=1^*$
theory with simply laced groups has been studied in series to
fourth order in the glueball superfield \cite{Aganagic:2003xq}.
Their method is the matrix model calculation developed by
Dijkgraaf and Vafa \cite{Dijkgraaf:2002fc}. This glueball
superpotential has been checked for the $\CN=1^*$ theory with the
$A_r=SU(r+1)$ gauge theory. However we find a small discrepancy
with the result in Ref.\cite{Aganagic:2003xq} at the fourth order
for other simply laced groups, like $D_r, E_6,E_7,E_8$.} We find
the glueball superpotential in a series, by first doing weak
coupling expansion of our superpotential and then taking the dual
or mirror transformation. We find that the glueball
superpotentials obtained from our effective superpotentials for
periodic and twisted boundary conditions are identical to each
other in a few leading orders.

The plan of this paper is as follows. In Sec.2, we introduce the
$\CN=1^*$ theory on $R^{1+2} \times S^1$ with periodic and twisted
boundary conditions. We focus here on the possible large gauge
transformations and the range of the Wilson-loop variables. In
Sec.3, we study the monopole instantons, or fractional instantons,
classically and quantum mechanically, which leads to the
preliminary version of the superpotential. In Sec.4, we present
the M-theory picture of the $\CN=4$ theory with various classical
groups and boundary conditions, to see that all twisted cases are
self-dual under $SL(2,Z)$. In Sec.5, we impose the $SL(2,Z)$
modular property for the twisted case, which fixes the
superpotential for a given group uniquely. In Sec.6, we obtain the
glueball superpotentials in weak coupling series by the dual
transformation and show that they are independent of the boundary
condition. In Sec. 7, we conclude with some remarks. In the
appendix, we also record glueball superpotentials for (untwisted)
theories associated with non-simply-laced algebra, as well as
provide various mathematical facts on Lie Algebra, affine Lie
algebra, and twisted affine Lie algebra.

\section{Theory}

The theories we consider are $\CN=1$ and $\CN=1^*$ supersymmetric
Yang-Mills theory with a simple Lie algebra $\CG$ of rank $r$. The
gauge field is defined as $A_\mu= A_\mu^i T^i$ with the
orthonormal elements $T^i$ of Lie algebra satisfying the inner
products $\Tr \;T^i T^j=\delta^{ij}$ as given in Appendix B. We
choose the Euclidean action for the gauge field to be
\be S_E = \frac{1}{2e^2} \int d^4x \; \Tr(F_{\mu\nu}F^{\mu\nu}) -
\frac{i\theta}{16\pi^2}\int d^4x\; \Tr (
F_{\mu\nu}\tilde{F}^{\mu\nu}) \,.\ee
The $\CN=1$ supersymmetric Yang-Mills theory has an additional
term for the gaugino field $\lambda$. With the holomorphic
coupling constant
\be \tau = \frac{\theta}{2\pi} + \frac{4\pi i}{e^2} \,,\ee
and the glueball superfield of the chiral gauge field strength
$W_\alpha$,
\be S= \frac{1}{16\pi ^2} \Tr(W_\alpha W^\alpha)\,, \ee
the Minkowski Lagrangian for the $\CN=1$ supersymmetric theory is
\be L = \bigl. 2\pi i\tau S \bigr|_{\theta^2} +{\rm h.c.} \ee

The $\CN=1^*$ theory has three more chiral fields $\Phi_i,
i=1,2,3$ in the adjoint representation with the standard kinetic
term and an additional $F$ term,
\be W = \Tr\left((\Phi_1,[\Phi_2,\Phi_3])+\sum_{i=1}^3m_i\Phi_i^2
\right) \,.\ee
When all $m_i$ vanish, the theory becomes the maximally
supersymmetric $\CN=4$ theory. When one of them vanishes and the
rest two become identical, the theory becomes the $\CN=2^*$
theory.

Classically, the vacuum of the $\CN=1$ theory is trivial and
unique. The $\CN=1^*$ theory has degenerate classical vacua,
\be <\Phi_i> = \sqrt{\frac{m_1m_2m_3}{m_i}} J_i
\label{classv}\,,\ee
where $J_i$'s are the $SU(2)$ generators embedded in the Lie
algebra $\CG$ \cite{Vafa:1994tf}.

Quantum mechanically, physics is much richer. Perturbatively the
$\CN=1$ theory is asymptotically free in high energy with a
confinement scale $\Lambda$. The vacuum of the $\CN=1$ theory is
strongly interacting and confining with a gaugino condensation.
There exist degenerate supersymmetric vacua. The vacuum physics
can be elegantly summarized by the Veneziano-Yankielowicz
potential \cite{Veneziano:1982ah} for the glueball field $S$,
which is a holomorphic function of the chiral coupling constant
$\tau$ and the chiral glueball superfield $S$ such that
\be{\cal W}^{VY}_{G} = hS -S \ln \left(
\frac{S}{\Lambda^3}\right)^h \label{VYpot}\,, \ee
where
\be \Lambda^3 = m^3 e^{2\pi i \tau(m)/(3h)} \prod_{a=0}^r\left[
\frac{k_a^*\balpha_a^2}{2}\right]^{k_a^*}\,, \ee
with the simple roots $\balpha_a$ of the extended Dynkin diagram
for the gauge Lie algebra and the comarks $k_a^*$ which is
discussed in Appendix B. The parameter $\Lambda$ is the
confinement energy scale found explicitly in
Ref.~\cite{Davies:2000nw} and the dual Coxeter number $h$ is
defined as the sum of comarks $h=\sum_{a=0}^r k_a^*$.

Here we consider the $\CN=1$ theory in the $\CN=1^*$ context which
is ultra-violet finite and the coupling $\tau(m)$ is defined on
the mass scale $m=m_1=m_2=m_3$. A vacuum is one of the stationary
points of this superpotential. The number of degenerate vacua is
$h$ and each vacuum is characterized by an integer $k$ such that
$0\le l \le h-1$. The glueball expectation value at a vacuum is
then
\be <S> = \frac{1}{16\pi^2} < \Tr \lambda \lambda > = \Lambda^3
e^{\frac{2\pi i l}{h}} , \;\; l=0,1,...h-1 \label{glueball1}\,,
\ee
where ${\cal W}_{VY}$ is stationary.

The $\CN=4$ theory with the $SU(N)$ gauge group is finite
perturbatively, and has the nontrivial strong-weak or $SL(2,Z)$
modular symmetry under which the coupling constant transforms as
\be \tau \rightarrow \frac{a\tau +b}{c\tau+d} \label{sl2z1}\,,\ee
for integers $a,b,c,d$ such that $ad-bc=1$. The $\CN=1^*$ theory
is finite at energy scale larger than mass scales $m_i$. For the
semiclassical calculation, we are initially interested in the weak
coupling limit where $e^2/4\pi^2$ is small at high energy. The low
energy physics of the $\CN=1^*$ theory would depend on which
classical vacuum one resides. If there is an unbroken nonabelian
subgroup, its coupling starts to grow and its matters become
confining at the energy scale much smaller than $m_i$.

The classical vacuum (\ref{classv}) of the $\CN=1^*$ theory has a
rich quantum structure as any unbroken nonabelian subgroup runs to
the confining phase, inducing also gluino condensation and
degenerate vacua. When the embedding is maximal, the gauge
symmetry is completely broken at the mass scale $m$ and all
particles become massive. When the embedding is trivial, the gauge
symmetry is not broken at all and so one ends up with the
confining phase. When the embedding is nontrivial but not maximal,
the low energy physics can depend on symmetry breaking pattern. If
the $SU(2)$ embedding has only  irreducible pieces of identical
dimension, only unbroken symmetry is a nonabelian subgroup and so
the theory reduces to the confining phase of smaller nonabelian
subgroup. If the embedding has irreducible pieces of different
dimensions, there would be an unbroken abelian symmetry whose
photon remains massless in the infrared. For the theory with
$SU(N)$ gauge group, all massive vacua, with no massless modes,
transform each other under the $SL(2,Z)$ transformation,
regardless whether they are in Higgs phase or confining
phase\cite{Donagi:1995cf}.

\subsection{Twisted Boundary Condition on $S^1$}

We compactify one of the spatial directions along a circle of
radius $R$, so that the fourth coordinate is cyclic,
\be x^4 \sim x^4+ 2\pi R \,.\ee
One could impose periodic boundary condition on the gauge field,
\be A_\mu(x^0,x^1,x^2,x^4+ 2\pi R) = A_\mu(x^0,x^1,x^2,x^4)\,, \ee
as well as other fields, which does not violate supersymmetry.
Thus we can use the holomorphic properties of the effective
superpotential in the compactified theories.

For a Lie algebra $\CG$ whose Dynkin diagram has a symmetry, there
exists an outer automorphism $\bsigma$ on the Lie algebra as
argued in the Appendix D such that $\bsigma^L=1$. We choose
twisted boundary condition on all fields so that the gauge field
satisfies
\be A_\mu(x^0,x^1,x^2,x^4+ 2\pi R) =
\bsigma(A_\mu(x^0,x^1,x^2,x^4))\,, \ee
with the same boundary condition on all other fields. This twisted
boundary condition also preserves supersymmetry.

For periodic boundary condition, we may take $L=1$. When there is
a nontrivial outer automorphism for Lie algebra $\CG$ so that
$L\ne 1$, the elements of the Lie algebra ${\cal G}$, as shown in
Appendix D, can be classified to the sets, ${\cal G}_n$'s, of the
eigenstates of the automorphism as $\bsigma({\cal G}_n) = e^{2\pi
i n/L} {\cal G}_n$, where $n=0,1,... L-1$. Since
$[\CG_m,\CG_l]=\CG_{m+l\; {\rm mod} \; L}$, only $\CG_0$ is a Lie
algebra and other sets would belong to representations of $\CG_0$.
For the Lie algebra $A_r, D_r, E_6$, an outer automorphism with
$L=2$ is allowed. For $D_4$, one can also have an outer
automorphism of $L=3$.

The gauge fields in the Lie algebra $\CG$ can be expressed in
terms of $\CG_n$ basis. Those that are associated with the
generators ${\cal G}_n$ can be Fourier-expanded as
\be A_\mu^i(x^4) T^i = \sum_{m=-\infty}^\infty\sum_{n=0}^{L-1}
A_\mu^{i,(m,n)} e^{-\frac{ix^4}{R}(m+\frac{n}{L} )} T^i =
\sum_{m=-\infty}^\infty\sum_{n=0}^{L-1} A_\mu^{i,(m,n)}
T^i_{m+\frac{n}{L}}\,, \ee
defining
\be T^i_{m+\frac{n}{L}} \equiv
e^{-\frac{ix^4}{R}(m+\frac{n}{L})}\; T^i , \;\; {\rm when } \;\;
T^i \in {\cal G}_n\,. \ee
Then the set $\{ T_{m+\frac{n}{L}}^i \}$ of these generators form
a twisted affine Lie algebra $\CG^{(L)}$.

The gauge fields without the $x^4$ dependent component would have
no Kaluza-Klein mass term. Thus with twisted boundary condition,
the gauge algebra $\CG$ of rank $r$ in the three dimensional limit
is reduced to a smaller simple Lie algebra $\CG_0$ of rank $r'<r$.
The Cartan subalgebra of $\CG_0$ is of dimension $r'$. Since
$[\CG_0,\CG_n]\in\CG_n$, $\CG_n$ belongs to a representation of
$\CG_0$ and every element of the ${\cal G}_n$ defines a weight of
the Lie algebra ${\cal G}_0$. The simple roots
$\bbeta_a,a=1,2,...r'$ of $\CG_0$ and the lowest negative weight
$\bbeta_0$ of $\CG_1$ define the Dynkin diagram of the twisted
affine Lie algebra $\CG^{(L)}$, which is shown in Figure 3 in
Appendix E.

In the twisted affine algebra $\CG^{(L)}$, each root vector
$\bbeta$ of $\CG_0$ is extended to a root $(\bbeta,m) $ with an
integer $m$ with the corresponding step operator $E_\bbeta^m$, and
each weight vector $\bbeta$ of $\CG_n$ is extended to a root
$(\bbeta, m+ \frac{n}{L})$ with the corresponding step operator
$E_\bbeta^{m+ \frac{n}{L}}$. Without the central term in the
twisted affine algebra like the case studied here, the square size
of these root is just given by $\bbeta^2$ regardless of the degree
$d=m+n/L$.

For the gauge theory with twisted boundary condition, the allowed
gauge transformation $g(x^i,x^4)$ should preserve the twisted
affine Lie algebra relations. There are small gauge
transformations which are generated by $\CG_0$ and independent of
$x^4$. There are also large gauge transformations with $x^4$
dependence. They would play a crucial role in determining the
classical vacuum structure of the theory as we will see.

\subsection{Wilson-loop Symmetry Breaking}

We focus on the Coulomb phase where classically the vacuum
expectation value of the scalar fields $\Phi_i$ vanishes. On a
circle, the fourth component gauge field can take a constant
nonvanishing expectation value in the Cartan subalgebra,
\be \langle A_4 \rangle = \bxi \cdot \bH \,,\ee
with $\bsigma(\bxi)=\bxi$ and so it belongs to $\CG_0$. This
configuration has zero field strength and so is a classical vacuum
configuration. For the generic expectation value of $\bxi$, the
gauge symmetry would be broken to its maximal abelian subgroup
$U(1)^{r'}$. The rest of the gauge components would get the mass
terms due to the Higgs mechanism and the Kaluza-Klein mass.

When the gauge symmetry is maximally broken to its abelian
subgroup, we introduce a dimensionless scalar field $\bvarphi$ of
$r'$ components such that
\be \bvarphi\cdot \bH = 2\pi R A_4 \,,\ee
and so $\langle \bvarphi \rangle = 2\pi R \bxi $. The Wilson loop
expectation value in the classical vacuum would be $\langle
P\exp(i\int_0^{2\pi R} dx^4 A_4) \rangle= \exp(i\bvarphi\cdot
\bH)\,.$

The allowed large gauge transformations are those which leave the
twisted affine algebra relations invariant. Easier ones are those
that transform even degree generators to even degree ones and odd
degree ones to odd degree ones. On the other hand, it is possible
to have large gauge transformations which would transform even
degree generators to odd ones and vice versa. Let us try the
general form of an allowed large gauge transformation as
\be U = e^{\frac{-ix^4}{R} \bw^*\cdot \bH} \label{wallowed}\,,\ee
where $\bw^* = \sum_{a=1}^{r'} c_a \bw_a^*$ with $\bw_a^*$ being
the fundamental coweight vector so that $\bw_a^* \cdot
\bbeta_b=\delta_{ab}$. Constraints on the coefficient $c_i$ arise
as we require that the gauge transformation of any step operator
\be U E_{m + \frac{n}{L}}^\bbeta U^\dagger = e^{-\frac{i
x^4}{R}\bw^*\cdot \beta} E_{m + \frac{n}{L}}^\bbeta \;\,, \ee
should be a step operator with root $(\beta, m+ \frac{n}{L} +
\bw^*\cdot \bbeta)$. This is easy to work out for all twisted
affine algebra by studying each case. For $A_{2r}^{(2)}$ the
allowed ones are given by the integer lattice defined by the set
of one half $\bw_a^*/2$ of each fundamental co-weight. For the
rest of the affine algebra the allowed ones are given by the
weight lattice $\Lambda_w$ generated by the fundamental weights
$\bw_a$ not by coweights $\bw_a^*$, contrasted to the theories
with non simply-laced untwisted affine Lie algebra
\cite{Davies:2000nw}.

These allowed large gauge transformations can be used to put the
vacuum expectation value to lie in a fundamental cell. For the
theory with $A_{2r'}^{(2)}$ the fundamental cell is
\be 0 \le \langle \bvarphi \rangle \cdot \bbeta_a < \pi, \;\;
a=1,...r' \label{ranget1}\,. \ee
For the theories with the rest of twisted affine algebra, the
fundamental cell is
\be 0 \le \langle \bvarphi \rangle \cdot \bbeta_a^* < 2\pi, \;\;
a=1,...r' \label{ranget2}\,,\ee
where $\bbeta_a$ is the simple roots of ${\cal G}_0$. This
periodicity of the vacuum expectation value will be manifest in
the supersymmeric effective potential obtained later. For the
reference, we enlist the vacuum expectation value for the theory
with periodic boundary condition also,
\be 0\le \langle \bvarphi \rangle \cdot \balpha_a< 2\pi, \;\;
a=1,...r \,,\ee
where $\balpha_a$ are the simple roots of $\CG$. This is again
consistent with the effective superpotential.

When nontrivial Wilson-loop breaks the gauge symmetry $\CG_0$ to
abelian subgroup $U(1)^{r'}$, besides large gauge transformation,
there are still discrete nonabelian gauge transformations, which
are Weyl-reflections on the root vectors of $\CG_0$. This reduces
the range of the vacuum expectation value. Especially, one can
always find Weyl reflections to reduce the range of the
expectation value in the fundamental cell to
\be \frac{4\pi}{L\bbeta_0^2} + \langle \bvarphi \rangle \cdot
\bbeta_0^* \ge 0 \label{alpha0}\,. \ee
This condition reduces the fundamental cell to a smaller subset,
which we may call the fundamental alcove. This reduced vacuum
moduli space can be characterized by a quotient space,
\be \bvarphi \in {\cal M}_{vacuum} = \frac{ R^{r'} }{2\pi\cdot
\Lambda_{LG} \rtimes W_{\CG_0}} \label{rangephi}\,, \ee
where $\Lambda_{LG}$ is the lattice of the allowed large gauge
transformations made by $\bw^*$, and $W_{\CG_0}$ is the Weyl group
on $\CG_0$.

The low energy dynamics of maximally broken theory would be
$x^4$-independent and purely abelian with the Euclidean action,
\be S_1 = \frac{2\pi R}{e^2}\int d^3 x \left(
\frac{1}{4\pi^2R^2}(\partial_i \bvarphi)^2 + {\bf F}_{ij}^2
\right) - \frac{2\pi i\theta R}{8\pi^2}\int d^3x
\epsilon_{ijk}\partial_i \bvarphi\cdot {\bf F}_{jk}\,, \ee
where ${\bf F}_{mn}$ is the purely abelian three dimensional gauge
field strength. One can dualize the three dimensional abelian
gauge field by introducing a dimensionless scalar field $\bsigma$,
adding a topological action
\be S_{2} = -\frac{i}{4\pi} \int d^3x \epsilon_{ijk} \partial_i
\bsigma \cdot {\bf F}_{jk} \label{2action} \,,\ee
and regarding $\bsigma$ and ${\bf F}_{mn}$ as dynamical variables.
Integrating over $\bsigma$ imposes that ${\bf F}_{mn}$ should be a
gauge field strength. Integrating over ${\bf F}_{mn}$ and
introducing a complex scalar field
\be \bz= i(\tau \bvarphi + \bsigma) \label{zscalar}\,,\ee
lead to a simplification of the total action $S_1+S_2$ to the
action
\be S_{classical} = \frac{1}{8\pi^2 R} \int d^3x\; \frac{1}{{\rm
Im} \tau} \;\partial_i \bz^\dagger \partial_i \bz \,.\ee

Including a gaugino field leads to a chiral superfield $\bX$ in
four dimensional sense whose bosonic field is the complex scalar
field $\bz$. We consider the case where the compactification
radius $R$ is much smaller than the confinement scale. In other
words we consider the case where $\Lambda \ll R^{-1} \ll m$. The
initial coupling constant $e^2(m)$ is small and so is
$e^2(R^{-1})$. Thus we are calculating in the weak coupling limit.
After integrating out the massive modes of mass scale $1/R$ or
more, we end up with a low energy effective action
\be S_{eff} = 2\pi R \int d^3 x \left({\cal
K}(X,X^\dagger)|_{\theta^2\bar{\theta}^2} + {\cal
W}(X)|_{\theta^2} + {\cal W}(X^\dagger)|_{\bar{\theta}^2} \right)
\,,\ee
in leading order in derivative expansion. The holomorphic
quantities in the theory are the chiral fields $\bX$, the coupling
$\tau(m)$ defined at the scale $m$ and the mass parameter $m$.
(Here we put all the mass scales of the $\CN=1^*$ theory to be
equal, $m_i=m$, for simplicity.) The holomorphic part of the
effective action would depend only on these chiral quantities,
\be {\cal W}_{eff} = {\cal W}(\bX,m,\tau) \label{effpot1}\,. \ee

\section{Monopoles and Superpotential}

We are interested in the effective action near the confining
vacuum where $<\Phi_i>=0$, and so all $\Phi_i$ supermultiplets
have mass $m$. In the low energy effective action we are looking
for, there is no perturbative correction to the $F$-term or the
holomorphic superpotential. Thus, we are interested in
nonperturbative contributions to the effective superpotential.
This effective superpotential has been found for the $\CN=1$
theory with $SU(N)$ gauge group \cite{katzvafa,Davies:1999uw} and
other simple gauge groups \cite{Vafa:1998vs,Davies:2000nw} and
also for the $\CN=1^*$ theory with $SU(N)$ gauge group and
periodic boundary condition \cite{Dorey:1999sj}. The
nonperturbative effects are due to magnetic monopole instantons,
which can be regarded as fractional instantons. Such
configurations are possible as the classical vacuum of the abelian
gauge theory has degeneracy due to the magnetic flux.

The natural nonperturbative contributions come from instanton
configurations in four dimensional Euclidean spacetime $R^3\times
S^1$, which satisfy self-dual equations
\be B_i \equiv \frac{1}{2}\epsilon_{ijk}F_{jk} = F_{i4}\,, \ee
where $i=0,1,2$. As one spatial direction $x^4$ is compactified to
a circle, we can have Wilson-loop gauge symmetry breaking. Let us
assume a maximal symmetry breaking to the abelian subgroups. In
this case it turns out that the solutions of the self-dual
equations are made of more than just instantons. There are
magnetic monopoles which can be regarded as fractional instantons.

\subsection{Classical Solution}

These magnetic monopoles and instanton solutions have been
extensively studied before. Initially W. Nahm studied calorons in
the context of ADHM contruction for instantons and Nahm's equation
for magnetic monopoles \cite{Nahm:1983sv}. When there is no
Wilson-loop symmetry breaking, there is a single caloron or
instanton solution found by periodic array. In the string theory
context, the relation between instantons and magnetic monopole are
realized as D0 branes on D4 branes wrapping on $R^3\times S^1$ are
T-dualized to D3 brane on the dual circle warpped by D1 branes,
where some class of solutions are found with Wilson loop symmetry
breaking\cite{Lee:1997vp}. Explicit solution for a single
instanton in the SU(2) gauge theory has been found when the gauge
symmetry is broken to $U(1)$ by the Wilson loop symmetry
breaking\cite{Lee:1998bb}. In Ref.\cite{Kraan:1998kp}, the ADHM
contruction was T-dualized to the Nahm construction with
nontrivial Wilson loop symmetry breaking and caloron solutions are
explored. For the theory with periodic boundary condition and any
simple gauge group of rank $r$ which is maximally broken by the
Wilson loop, a single instanton is made of $r+1$ constituent
fundamental monopoles \cite{Lee:1998vu}. For the theories with
twisted boundary condition, some aspects of Wilson loop symmetry
breaking and instantons are studied \cite{Hanany:2001iy}.

Here we provide enough detail for the sake of completeness. Let us
first write down a single magnetic monopole solution in the
$SU(2)$ gauge theory. With the asymptotic value of the Higgs field
$u$, the BPS solution is given as
\be V^a_i({\bf r},u) = \epsilon_{aij}\hat{r}^j\left(\frac{1}{r} -
\frac{u}{\sinh ur} \right) , \;\;\;\; \Phi^a({\bf r},u) =
\hat{r}^a\left( \frac{1}{r}-u\coth ur\right) \label{pssol} \,,\ee
which satisfies the BPS equation $\partial_i \Phi^a +
\epsilon^{abc} V^b_i \Phi^c = \epsilon_{ijk}( \partial_jV^a_k +
\epsilon^{abc} V^b_j V^c_k/2) $.

For our theory with twisted boundary condition, we can choose any
root $(\bbeta, m + \frac{n}{L})$ of degree $m+n/L $ of the twisted
affine algebra $\CG^{(L)}$ and its corresponding step operator
$E^{m+n/L}_\bbeta$. There are corresponding $SU(2)$ generators,
\be t^1 = \frac{1}{2}( E_\bbeta^{m+n/L}+ E_{-\bbeta}^{-m-n/L}),
\;\; t^2 = \frac{1}{2i}( E_\bbeta^{m+n/L}- E_{-\bbeta}^{-m-n/L}),
\;\; t^3=\frac{1}{2}\bbeta^*\cdot \bH \,.\ee
Note that the exchange of two roots $(\bbeta, m+s)$ and
$(-\bbeta,-m-s)$ in the above definition is just a $\pi$-rotation
along $t^3$ direction. We can embed the monopole solution
(\ref{pssol}) with these SU(2) generators. First we note that \be
t^a(x^4) = e^{-i(m+n/L)x^4t^3/R} \tilde{t}^a e^{+i(m+n/L)x^4
t^3/R}\,, \ee for $x^4$-independent $\tilde{t}^a$. Here the
solution of the self-dual equation $B_i = F_{i4}$ is then
\be A_i = V^a_i({\bf r}, u) t^a , \;\;\; A_4 = (\bxi-\frac{1}{2}
(\bxi\cdot \bbeta)\bbeta^*) \cdot \bH + \Phi^a({\bf r},u) t^a
\,,\ee
where
\be u = \bxi \cdot \bbeta + \frac{1}{R}(m + \frac{n}{L}) \,.\ee
The Euclidean action for the magnetic monopole is then
\be S_E = \frac{4\pi}{e^2} \frac{2}{\bbeta^2} 2\pi R\left| \bxi
\cdot \bbeta + \frac{1}{R} ( m + \frac{n}{L}) \right| \,.\ee
(However $e^{-i(m+n/L)x^4t^3/R}$ is not in general an allowed
large gauge transformation and so the monopole would be
intrinsically $x^4$ dependent \cite{Lee:1998vu,Hanany:2001iy}.)

The detailed zero mode analysis of these monopole solutions would
lead to the constituent monopole structure. Those with only four
zero modes, three for its position and one for the phase, would be
called fundamental monopoles. The rest of the monopole
configurations would be composite of these fundamental monopoles.
In this analysis\cite{Lee:1998vu,Hanany:2001iy}, it is crucial to
transform the $A_4$ expectation values by  large gauge
transformations and by Weyl reflections so that it lies in the
fundamental alcove defined by
Eqs.~(\ref{ranget1},\ref{ranget2},\ref{alpha0}).

The solutions with only four zero modes are those for the simple
roots $(\bbeta_a,0), a=1,..r'$ and the lowest root $(\bbeta_0,1)$.
(Of course the solutions for negative of the above roots appearing
in the Dynkin diagram is just a gauge transformation of the
original one.) The Euclidean action for each fundamental monopole
becomes
\be S_0= \frac{8\pi^2}{e^2} \left(\frac{2}{L\bbeta_0^2}
+\frac{1}{2\pi}\langle \bvarphi \rangle \cdot \bbeta_0^* \right),
\;\; S_a= \frac{8\pi^2}{e^2} \frac{1}{2\pi} \langle\bvarphi\rangle
\cdot \bbeta_a^* , a=1,2,...r' \label{monoact1}\,.\ee
In the fundamental alcove we consider, the above actions are all
positive. Let us consider a configuration which is made of
$\tilde{k}_a^*, a=0,1,...r'$ number of $\bbeta_a$ fundamental
monopoles where $\tilde{k}_a^*$ are the comarks of the twisted
affine algebra $\hat{\CG}^{(L)}$ considered in Appendices B and E
and shown explicitly at Figure 3 of Appendix E. We assume that it
is possible to superpose self-dual magnetic monopole
configurations to obtain another self-dual configurations, which
seems reasonable. As the comarks $\tilde{k}_a^*$ satisfy the
following relations,
\be \sum_{a=0}^{r'} \tilde{k}_a^* \bbeta_a^* = 0\,, \;\;\;
h=\sum_{a=0}^{r'}\tilde{k}_a^*\,, \;\;\;\;
\frac{2\tilde{k}_0^*}{L\bbeta_0^2} = 1 \,,\ee
with the dual Coxeter number $h$, the above composite
configuration would have $4h$ zero modes, zero magnetic charge and
the action of value $8\pi^2/e^2$. This is exactly the property of
a single instanton configuration on $R^3\times S^1$. Thus one can
regard instantons as being made of constituent magnetic monopoles,
which carry fractional instanton charges given by their action.

Any general self-dual configurations would be characterized by a
$r'+1$ non-negative integer set $\{ n_a, a=0,1,...r'\}$, which
count the number $n_a$ of the fundamental monopole $\bbeta_a^*$.
The total magnetic charge ${\bf g}$ and action $S$ of such
configuration would be
\be {\bf g}= \sum_{a=0}^{r'} n_a \bbeta_a^* \,,
\;\; S= \sum_{a=0}^{r'} n_a S_a \label{sume}\,, \\
\ee
where $S_a, \; a=0,1,...r'$ are those in Eq.~(\ref{monoact1}). The
number of the zero modes would be $4\sum_{a=0}^{r'} n_a$.

\subsection{Superpotential: Dilute Instanton Gas}

Let us consider the quantum effect of the magnetic monopole
instantons on $R^{1+2}\times S^1$. Classically a theory with
$U(1)^{r'}$ gauge group has degenerate vacua whose magnetic flux
can take arbitrary value. The magnetic monopole instantons tunnel
between these degenerate vacua. When we consider the long distance
physics whose length scale is much larger than the
compactification radius $R$, or the magnetic monopole size scale,
the magnetic monopole instantons induce an effective action.

To find this effective action, we reconsider the action
(\ref{2action}) for the dual photon by regarding ${\bf F}_{jk}$ as
a field strength and write the action as a boundary value,
\be S_2 = -\frac{i}{2\pi }\int_{S^2} dx^m\;\; \bsigma \cdot {\bf
B}_m \label{22action} \,,\ee
where ${\bf B}_i = \epsilon_{ijk}{\bf F}_{jk} /2$. For a magnetic
monopole of magnetic charge ${\bf g}$, asymptotically ${\bf B}_m =
({\bf g}\cdot \bH/2 ) x^m/x^3$ and the above action becomes $S_2 =
-i\langle \bsigma \rangle \cdot {\bf g}$. Only place where the
$\bsigma$ expectation value appears is the exponential $e^{-S_2}$
of the monopole effective action. Since the magnetic charge ${\bf
g}$ is quantized in $\bbeta_a^*, a=0,1,2...,r'$, the scalar field
$\bsigma$ would be physically equivalent to $\bsigma+2\pi \bw_a,
a=0,2...r'\,,$ where $\bw_0$ is the fundamental weight vector for
the root $\bbeta_0$. By going case by case, one can show that the
fundamental cell of $\bsigma$ for the case with twisted boundary
condition is identical to that of $\bvarphi$ as given in
Eqs.~(\ref{ranget1}) and (\ref{ranget2}). For the theory with non
simply laced group, the $\bvarphi$ and $\bsigma$ have the
different ranges~\cite{Davies:2000nw}. The general range of
$\bvarphi$ in Eq.~(\ref{rangephi}) is translated to the range for
the complex vector field $\bz=i(\tau \bvarphi +\bsigma)$ as
\be \bz \in {\cal M}_{\rm vacuum} = \frac{C^{r'}}{
\Lambda^c\rtimes W_{\CG_0} } \label{zrange}\,,\ee
where $\Lambda^c $ is a lattice defined by $2\pi i( \tau \bw +
\bw')$ where $\bw, \bw'$ are vectors appeared in the allowed large
gauge transformations (\ref{wallowed}). For the twisted Lie
algebra with classical groups, the complex $\bz$ denotes the
positions of D2 branes in the dual torus in the M-theory as we
will see in the next section.

Including the dual photon contribution $S_2$ of
Eq.~(\ref{22action}) and the topological term
$\frac{-i\theta}{4\pi^2}\int_{S^2} dx^m \bvarphi {\bf B}_m$, the
Euclidean action for each fundamental monopole can be calculated
for a given asymptotic value of the complex field $\bz = i(\tau
\bvarphi+\bsigma)$. Their actions (\ref{monoact1}) become
generalized to holomorphic ones,
\be - S_0 = \frac{4\pi i \tau}{L\bbeta_0^2 } + \bbeta_0^*\cdot
\langle \bz \rangle , \;\;\; -S_a= \bbeta_i^* \langle \bz \rangle
\label{monoact2} \,,\ee
which give the exponential suppressing factors as the real value
of the negative action is negative.

In the small radius limit, where $m^{-1}\ll R \ll
\Lambda_{CDR}^{-1} $, one can in principle calculate the
contribution to the superpotential by fundamental monopoles by
calculating the massless gluino correlation functions. Following
the argument in Ref.\cite{Davies:1999uw}, we choose the running
coupling to be given at the mass scale $m$, and replace the
expectation value $\langle \bz \rangle$ by the dynamical chiral
field $\bX$. As we consider the $\CN=1^*$ theory, which is finite
at high energy, the regularization should be somewhat different
from that of the Pauli-Villars regularization. The calculation
would be more or less identical to that in
Ref.\cite{Davies:1999uw} and we obtain the leading contribution to
the effective potential (\ref{effpot1}) as
\be {\cal W}_{\CN=1}^{\rm twisted} = m^3 \left(
\frac{2}{\bbeta_0^2} e^{\frac{4\pi i \tau}{L\bbeta_0^2} +
\bbeta_0^*\cdot \bX} + \sum_{a=1}^{r'} \frac{2}{\bbeta_a^2}
e^{\bbeta_a^*\cdot \bX} \right) \label{etoda}\,. \ee
This is the holomorphic superpotential for the $\CN=1$
supersymmetric theory with twisted boundary condition. The above
superpotential is the affine Toda potential for the associated
affine algebra. It is independent of the compactification radius
$R$ and so should be true for any value of $R$ due to holomorphic
property. Following Ref.\cite{Davies:1999uw} one can calculate the
vacuum degeneracy and the gluino condensation at each vacua by
finding the stationary points of the above potential. In the
infinite radius limit, the physics should be identical whether one
uses periodic or twisted boundary condition. Indeed one can easily
show that this is true due to the identities enlisted in Appendix
E for roots and comarks of the untwisted and twisted affine
algebra. We will work out this physics in detail by using the
glueball superpotential in Sec.6, where the above affine Toda
potential will be mapped to the Veneziano-Yankielowicz
superpotential.

\subsection{Superpotential: Summing Over Higher Instantons}

For $\CN=1^*$ theory, all other magnetic monopoles also seem to
contribute to the superpotential \cite{Dorey:1999sj}. Noting that
the step operator for the root $(\beta, m+n/L)$ and that for the
root $(-\bbeta, -m-n/L)$ lead to the gauge equivalent monopole
solutions, we can restrict to the positive roots $\bbeta$ in
calculating the monopole instanton contribution. Their
contribution to the negative action $-S(m,n)$ with $m\in Z$ and
$n=0,1,...(L-1)$ can be divided into two groups as follows,
\bear && -S^+_{m,s} = 2\pi i \tau \frac{2m}{\bbeta^2} + \left(
2\pi i \tau \frac{2n}{\bbeta^2 L}+ \bbeta^*\cdot \langle \bz
\rangle \right) , \;\;
(m\ge 0,\; L-1\ge n\ge 0) \\
&& -S^-_{m,s} = 2\pi i \tau \frac{2m}{\bbeta^2} - \left( 2\pi i
\tau\frac{2n}{\bbeta^2 L} + \bbeta^*\cdot\langle \bz \rangle
\right),\;\; (m\ge 1, \; L-1\ge n\ge0) \eear
As in for $SU(r+1)$ theory studied in Ref.\cite{Dorey:1999sj}, we
also expect the contribution from the $k$ number of every
$(\bbeta,m+n/L)$ monopole instantons whose action would be $k$
times that of $(\bbeta,m+ n/L)$ monopole. Thus in the
semiclassical approximation, the most general possible holomorphic
superpotential generated by the $(\bbeta,m+n/L)$ monopoles with
all possible $m$ would be
\be \CW(\bbeta, n/L) = \sum_{k=1}^\infty a_k y^k +
\sum_{k=1}^\infty \sum_{m=1}^\infty (q^{\frac{2}{\beta^2}})^{km}
\left( b_{k,m} y^k+ c_{k,m} y^{-k} + d_{k,m} \right)\,, \ee
where
\be q = e^{2\pi i \tau} , \;\;\; y= e^{\frac{4\pi i \tau
n}{\bbeta^2 L} + \bbeta^*\cdot \bX}\,. \ee
The direct calculation of the exact coefficients of these
contribution would be a challenge. The $y$ independent
contributions would be pure instanton contributions.

Fortunately, for the $A_r=SU(r+1)$ gauge theory with periodic
boundary condition, the above contribution of all magnetic
monopoles seems to be summed to a Weierstrass elliptic function,
which is justified from other considerations \cite{Dorey:1999sj}.
Here we also assume that a similar replacement does the trick.
Thus, the above action for the root $(\bbeta,n/L)$ of $\CG_n$
becomes
\be \CW(\bbeta,n/L) \sim \wp\left(\bbeta^*\cdot \bX + 2\pi i \tau
\frac{2n}{\bbeta^2L} ; 2\pi i , 2\pi i \tau \frac{2}{\bbeta^2}
\right)\,. \ee
To fix the various terms, we have used the series expansion (F.4)
of the Weierstrass elliptic function in Appendix F. There will be
one such expression for all positive roots in the $\CG_{n}$. In
summing over the contribution for each positive root, we fix the
coefficient of the elliptic function, by considering the
contributions from single fundamental monopole instanton given in
Eq.~(\ref{etoda}). Here it is also useful to consider the series
expansion of the Weierstrass elliptic function in Appendix F. In
addition, we put the same coefficient to the roots of the same
length for Weyl reflection symmetries. Then, we obtain a
preliminary holomorphic superpotential (\ref{effpot1}) for the
$\CN=1^*$ theory on $R^{1+2}\times S^1$ with twisted boundary
condition as
\be \tilde{\CW}_{{\cal G}^{(L)}} = m^3 \sum_{n=0}^{L-1}
\sum_{\bbeta\in {\cal R}_+( {\CG_n})} \frac{2}{\bbeta^2}\;\;
\wp\left( \bbeta^*\cdot \bX+ 2\pi i \tau \frac{2n}{L\bbeta^2 }\; ;
2\pi i ,2\pi i \tau \frac{2}{\bbeta^2}\right) \,,\label{monopot}
\ee
where ${\cal R}_+(\CG_n)$ is the set of positive roots in $\CG_n$.
Note that this potential is even under $\bX\rightarrow -\bX$ due
to the summation and so the above sum could be extended to the sum
over all roots in $\CG_n$ once the factor half is included. As we
choose the same normalization for equal length roots, the Weyl
reflection symmetry is manifest. This contrasts to the effective
potential (\ref{etoda}) for the $\CN=1$ theory where the Weyl
reflection symmetry is already used to put $\bX$ in the
fundamental alcove.

Here we present for the sake of completeness the $\CN=1^*$
effective superpotential for the $\CN=1^*$ theory of a simple Lie
algebra $\CG$ with periodic boundary condition, which is
\be \CW_{\CG^{(1)}} = m^3 \sum_{\balpha\in {\cal R}_+(\CG)}
\frac{2}{\balpha^2} \;\; \wp \left(\balpha^*\cdot\bX; 2\pi i ,
2\pi i \tau \frac{2}{\balpha^2} \right) \,,\label{untpot} \ee
For $A_r= SU(r+1)$, we recover the Dorey's
result\cite{Dorey:1999sj}. This superpotential for the untwisted
group $\CG^{(1)}$ can be recognized to be related to the twisted
elliptic Calogero-Moser model for the Lie group $\CG$. They have
appeared as the integrable model related to the Seiberg-Witten
curve for the $\CN=2^*$ theory of the gauge algebra $\CG$
\cite{D'Hoker:1998yi}. Also they appear in the study of the
relation between the integrable model and the $\CN=1^*$
theory\cite{Kumar:2001iu}. Contrast to these works, our result has
definite normalization for the superpotential. This will allow the
study of the modular property of the effective superpotential in
more detail.

Let us now consider the three dimensional limit of $\CN=1^*$
theories as done for the $SU(r)$ case in Ref.~\cite{Dorey:1999sj}.
We take a small $R$ limit so that $R^{-1}> m^{-1}$, keeping the
three dimensional coupling constant $2\pi R/e^2(m)= g_3^2 $ fixed,
which means a small coupling or small $q$ limit. The corresponding
three dimensional effective potential would be $\CW_3= 2\pi R
\cdot\CW$. With periodic boundary condition, the three dimensional
gauge group would remain $\CG$ and all $q$-dependence disappears.
>From the formula in Appendix F, we can read
\be \CW_3(\CG)= 2\pi R m^3\sum_{\alpha\in{\cal R}_+(\CG)}
\frac{2}{\balpha^2} \frac{1}{\sinh ^2(\frac{\balpha\cdot X}{2}) }
\label{three} \quad.\ee
(Here to achieve the three dimensional limit one should rescale
the mass parameter to absorb the length scale $R$.) We expect the
three dimensional limit of the theories with twisted boundary
condition leads to the $\CN=1^*$ theory with gauge group $\CG_0$.
Indeed in this limit all instanton contribution from the twisted
sectors $\CG_n, n\ne 0$ disappear. The three dimensional limit of
the superpotential (\ref{untpot}) for the non simply laced group
$\CG$ becomes identical to the above expression (\ref{three}) once
the three dimensional algebra $\CG_0$ is identified with the non
simply laced Lie algebra $\CG$. This is a consistency check for
our effective superpotential.

\section{$SL(2,Z)$ Symmetry }

\subsection{$SL(2,Z)$ and Four Types of Orientifold Planes}

The $\CN=4$ supersymmetric Yang-Mills theory with $SU(r)$ symmetry
on $R^{1+3}$ dimension describes the low energy dynamics of $r$
parallel D3 branes of the type IIB theory when their mutual
distance is smaller than the string length. The type IIB string
theory has the $SL(2,Z)$ symmetry and $D3$ branes are invariant
under it. This symmetry manifest nonpertubratively in the $\CN=4$
supersymmetric gauge theory on $D3$ branes as a generalization of
weak-strong coupling duality.

To introduce the other classical groups on $R^{1+3}$, we need to
introduce the three-dimensional orientifold parallel to D3 branes.
(See Ref.~\cite{Dabholkar:1997zd} for a review.) There are four
kinds of them, $O3^-, O3^+, \widetilde{O3}^+, \widetilde{O3}^-$,
whose low energy field theory with $r$ parallel $D3$ branes has
gauge algebra, $D_r$, $C_r$, $C_r$ and $B_r$, respectively. Under
the $S$-duality two orientifolds $O3^-$ and $\widetilde{O3}^+$ are
invariant and two orientifolds $O3^+ $ and $\widetilde{O3}^-$ are
interchanged. Under the $T$ transformation of the $SL(2,Z)$, two
orientifolds $O3^+$ and $\widetilde{O3}^+$ are interchanged.

Perhaps not widely recognized, although elementary, is the nature
of two $Sp(r)$ theories out of orientifold construction. We know
that there exists only one $\CN=4$ $Sp(r)$ Yang-Mills theory in
four dimensions, so the question is what are the distinction
between the two $+$ type orientifolds in terms of the field
theory.

The answer lies in the $\theta$ angle. In string theory convention
in $3+1$ dimension, the long roots of the $C_r=Sp(r)$ theory have
length square four and so the range of $\theta $ is increased to
$4\pi$ instead of $2\pi$. On the other hand, the $\theta$ angle of
$\CN=4$ theory is inherited from the type IIB axion which has a
natural period $2\pi$. Thus, there is an apparent discrepancy of
the period between type IIB picture and Yang-Mills picture.

However, when we say IIB axion having period $2\pi$, it actually
means that $T$ transformation of IIB $SL(2,Z)$ shifts the axion by
$2\pi$. Since both $O3^+$ and $\widetilde{O3}^+$ has a discrete
$Z_2$ NS-NS 2-form flux and since $\widetilde{O3}^+$ has
additional RR 2-form flux, also $Z_2$ valued, the same $T$
transformation of IIB interchanges these two orientifolds and at
the same time shifts the axion by $2\pi$. From this, it is clear
how the $4\pi$ range of $\theta$-angle is generated from IIB
viewpoint. Half of this $4\pi$ range is covered by $O3^+$ and the
other half by $\widetilde{O3}^+$. There is no fundamental
distinction between the two cases, since by continuous deformation
of IIB axion combined with $T$ transformations, the two can be
mapped to each other.

\subsection{Compactification and M-theory Realization}

Another way to view the $SL(2,Z)$ symmetry of IIB theory and in
particular the $SL(2,Z)$ of four dimensional Yang-Mills theory is
to rely on M theory. The type IIB string theory on flat 10
dimension has the $SL(2,Z)$ symmetry. This symmetry can be
understood from the $M$ theory point of view by considering the
compactification of type IIB theory on a circle of radius $R$
along the $x^4$ coordinate, which is T-dual to the type IIA theory
on a dual circle of radius $1/R$. Type IIA theory on flat 10
dimensional space can be obtained from compactifying the M theory
on a circle along $x^{11}$ coordinate. Thus the M theory on two
torus $T^2$ is T-dual to the type IIB theory on a circle. The
obvious $SL(2,Z)$ symmetry of the torus compactification of the
$M$ theory is translated to the $SL(2,Z)$ symmetry of type IIB
theory on a circle \cite{Schwarz:1995dk}.

A flat M2 brane located on a point on $T^2$ would be a D2 brane on
type IIA theory, which is a $T$-dual of a D3 brane wrapped on the
circle. The M2 brane does not change under the SL(2,Z) of the
torus and so the wraped D3 branes is also a $SL(2,Z)$ singlet of
type IIB as we know well. We regard the circle along $x^4$ as the
compact circle of our spacetime $R^{1+2}\times S^1$. If we have
$r$ parallel D3 branes wrapping the circle along $x^4$ and lying
very close to each other in the transverse direction, the
effective low energy field would be the $\CN=4$ supersymmetric
field theory on $R^{1+2}\times S^1$ with gauge group $SU(r)$. The
expectation value $\langle \bsigma \rangle $ of the dual photon
shows the location of the $M2$ branes along $x^{11}$, and the
expectation value $ \langle \bvarphi \rangle $ shows the position
of $M2$ branes along the dual $\tilde{x}^4$ direction. As there is
nonzero $\theta$ parameter, two directions are not orthogonal but
skewed and $\langle \bz \rangle = \langle \tau\bvarphi+
\bsigma\rangle $ denotes the complex positions of $M2$ branes on
the two torus. In Figure 1, we draw the two torus $T^2$ along dual
$\tilde{x}^4$ and $x^{11}$ direction in the M-theory. When we take
the infinite radius limit, the dual two torus $T^2$ shrinks and so
the $SL(2,Z)$ symmetry will still hold.

We can add an O3 plane to D3 branes to get other classical gauge
groups in four dimension. The four kinds of O3 orientifolds lifted
to M theory has been fairly well discussed in
Ref.\cite{Hanany:2000fq} and here we summarize it briefly for the
completeness. Upon compactification, we wrap them along the $x^4$
circle. After T-duality a $O3$ plane breaks to two O2 planes on
the dual $\tilde{x}^4$ in the type IIA theory which makes the dual
circle to be a line segment. There are four kinds of O3
orientifolds as well as there are four kinds of two dimensional
planes $O2^-, O2^+, \widetilde{O2}^+, \widetilde{O2}^-$. Figure 1
shows that how four $O3$ planes are decomposed to two $O2$ planes
lying along the dual $\tilde{x}^4$ direction. From the M-theory
point of view a single two dimensional orientifold is made of two
OM2 planes lying along the $x^{11}$ direction. There are $OM2^+$
and $OM2^-$ planes. Thus a single $O3$ plane is made of four OM2
planes on a torus.

\begin{figure}[!t]
\hskip 2cm
\includegraphics[width=0.7\textwidth]{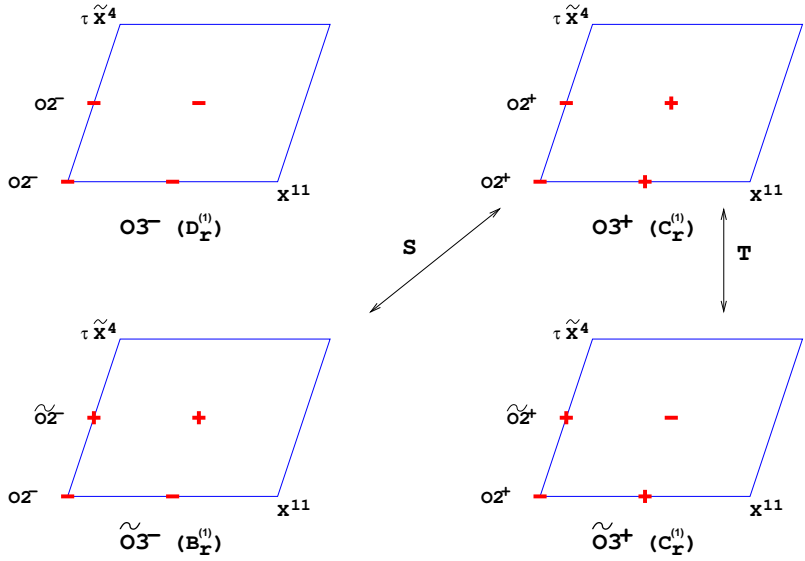}
\caption{The four diagrams show M theory realization of four types
of $O3$ planes compactified on a circle. After a T-duality, which
splits a single $O3$ to a pair of $O2$'s, we lift the
configurations to M theory by adding a $x^{11}$ circle. Each $O2$
is composed of two $OM2$'s, separated along $x^{11}$, of which
there are two types, $OM2^+$ and $OM2^-$. These are denoted by red
$+$'s and red $-$'s on the torus. The four dimensional limit
corresponds to shrinking the torus while maintaining its complex
modulus fixed.}
\end{figure}

Figure 1. also shows this M theory lift of the four kinds of O3
branes. The two signs denote the two possible OM planes. The
complex structure of $T^2$ becomes the coupling constant $\tau$.
The lattice of $OM$ planes on the $T^2$ could have some obvious
symmetries. The point to notice is that $O3^+$ and
$\widetilde{O3}^-$ are S-dual to each other, $\widetilde{O3}^+$
are self S-dual, and $O3^-$ are $SL(2,Z)$ symmetric, regardless of
the shape of $T^2$ or the $\tau$ coupling constant. In addition
the $T$ of the $SL(2,Z)$ transforms $O3^+$ to $\widetilde{O3}^+$.
>From the picture we can see the $SL(2,Z)$ of the $O3^-$
configuration or $D_r$ theory as expected. Also we can see the
$\Gamma_0(2)$ symmetry for the rest of $O3$ planes. This is the
$M$ theory picture of the $\Gamma_0(2)$ theory for the $\CN=4$
supersymmetric theories with $B_r,C_r$ gauge groups and periodic
boundary condition. As argued before these modular symmetry also
appear in the superpotential for the $\CN=1^*$ theory with
classical groups and periodic boundary condition.

\subsection{M Theory Realization of Twisted Theories}

Let us now consider the $\CN=4$ supersymmetric Yang-Mills theory
on $R^{1+2}\times S^1$ with twisted boundary condition. One can
read off the T-dual picture of the brane configuration. We need
two O2 planes and D2 branes along the dual circle, which can be
determined from the extended Dynkin diagram. The $A_{2r}^{(2)}$
theory arises from the low energy theory on one $\widetilde{O2}^-$
plane and one $O2^+$ or $\widetilde{O2}^+$ plane with D2 branes
inserted between them. The $A_{2r-1}^{(2)}$ theory arises from the
low energy theory on one $O2^-$ plane and one $O2^+$ or
$\widetilde{O2}^+$ plane. The $D_{r+1}^{(2)}$ theory arises from
the low energy theory on one $\widetilde{O2}^-$ plane and one
$\widetilde{O2}^-$ plane with D2 branes inserted between them.

\begin{figure}[!t]
\hskip 2cm
\includegraphics[width=0.7\textwidth]{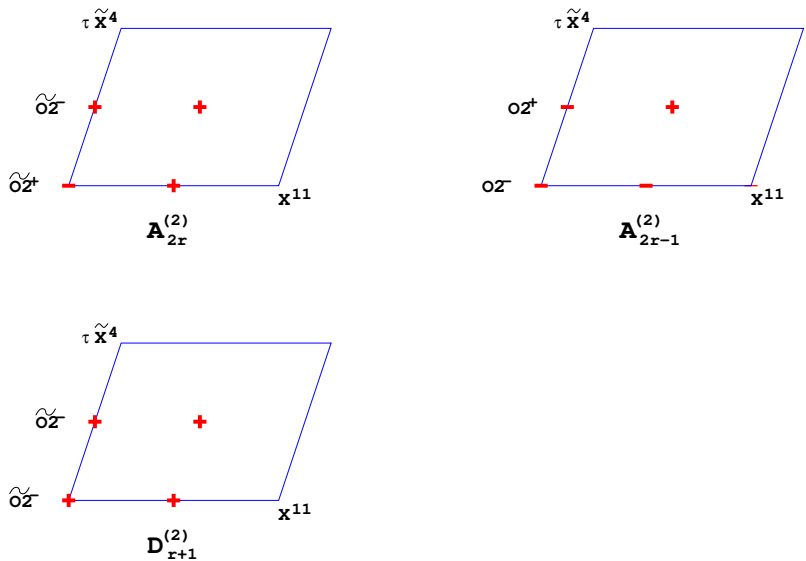}
\caption{M theory realization of $O2$ planes that gives the
twisted theories with classical groups upon T-duality. Again each
$O2$ is composed of two $OM2$'s, separated along $x^{11}$, of
which there are two types, $OM2^+$ and $OM2^-$, denoted by red
$+$'s and red $-$'s on the torus. Together with four possibilities
of Figure 1, these exhaust all possible such combination of
$OM2$'s up to overall translations of the torus.}
\end{figure}

Their corresponding M-theory pictures on $T^2$ are shown in Figure
2. One can see that the $ SL(2,Z)$ symmetry manifests for these
twisted case with arbitrary coupling constant $\tau$. Thus, the
corresponding $\CN=4$ theory with twisted boundary condition are
expected to have the exact $SL(2,Z)$ symmetry. Similar stringy
realization of the $\CN=4$ supersymmetric models in various
dimensions with twisted boundary condition has been realized in
Ref. \cite{Hanany:2001iy,Gimon:1998be}.

This fact that all twisted $(N=4)$ theories are invariant under
$SL(2,Z)$ may be quite surprising, since none of the extended
Dynkin diagrams in question are simply-laced. On the other hand,
note that all twisted theories arise from twisting a gauge theory
with simply-laced gauge groups, namely $A_r$, $D_r$, and $E_6$,
which, prior to the twisting, are all equipped with $SL(2,Z)$. The
twisting (upon compactification) must commute with the action of
$SL(2,Z)$ for four dimensional theories. We do not have a concrete
understanding of why this is so in the full string theory setting.

\section{$SL(2,Z)$ Symmetries and Exact Superpotential }

In this work we are searching for the superpotential for the
$\CN=1^*$ theory with twisted boundary condition, starting from
that for the $\CN=1$ theory. While one does not have exact control
over the multi-monopole instanton contributions, one may expect
that the resulting effective superpotential for each class of
magnetic charge would be expressible by Weierstrass elliptic
function as in the $SU(N)$ gauge group case, which was roughly
where we stopped in section 3. While the resulting superpotential
(\ref{monopot}) has the right weak coupling limit to the $\CN=1$
theory, there exists some ambiguity.

Our guess (\ref{monopot}) for the superpotential of the $\CN=1^*$
theory with twisted boundary condition has been shown to be
consistent with the Weyl reflection, the three dimensional limit,
and the week coupling limit or the affine-Toda limit. Hence one
may conclude our guess is the right one. However, this is too
hasty a conclusion. We will see our guess is almost but not quite
right. When we convert the sum of the monopole contributions to a
Weierstrass function, an ambiguity could arise from the
uncertainty of the pure instanton effect which depends only on
$q=e^{2\pi i \tau}$. This does not affect the above consistency
checks.

On the other hand, as we reviewed in the previous section, there
is a strong indication for the $SL(2,Z)$ duality of $\CN=4$
theories which would be inherited by $\CN=1^*$ theories. When we
take into account the $SL(2,Z)$ symmetry, the purely instantonic
ambiguity is resolved and we arrive at the somewhat modified new
conjecture. This new conjecture is unique and also passes an
additional consistency check given in the next section. In the
next section the weak coupling expansion of the glueball
superpotential will be obtained by the ``dual'' or ``mirror''
transformation. When the rank of the gauge group is small, the
pure instanton contributions appear quickly in a few leading terms
in the weak coupling series, and are playing a crucial role for
this consistency.

\vspace{3em}
\begin{center}
\begin{tabular}{|c|c|}
\hline gauge group and boundary condition & effective
superpotential \\ \hline $A_r^{(1)}, D_r^{(1)}, E_6^{(1)},
E_7^{(1)}, E_8^{(1)}$ & untwisted $A_r,D_r,E_6,E_7, E_8\;\;$ eCM
\\ \hline $B_r^{(1)}, C_r^{(1)}, G_2^{(1)}, F_4^{(1)} $ & twisted
$B_r,C_r,G_2,F_4 \;\;$ eCM
\\ \hline
$A_{2r}^{(2)} $ & twisted $BC_r$ eCM \\
$A_{2r-1}^{(2)}$ & untwisted $C_r$ eCM \\
$D_{r+1}^{(2)}$ & untwisted $B_r$ eCM \\
$E_6^{(2)} $ & untwisted $F_4$ eCM \\
$D_4^{(3)}$ & untwisted $G_2$ eCM \\
\hline
\end{tabular}
\end{center}

\begin{center} {\bf Table I}: The gauge group and boundary condition
of a $\CN=1^*$ theory on $R^{1+2}\times S^1$ and the corresponding
elliptic Calogero-Moser Model. \end{center}

\vskip 1cm

At the end of day, we are able to modify the above conjecture
(\ref{monopot}) of the superpotential in such a way that it
respects the $SL(2,Z)$ symmetry. The modification involves an
analytic function of $q=e^{2\pi i \tau}$ only, which can be
attributed to the pure instanton contributions to the
superpotential. Thus one could say the magnetic monopole
contribution written in the Weierstrass function as in the
previous section is fine but there was an ambiguity in the pure
instanton contribution, which will be fixed by the $SL(2,Z)$
symmetry. The resulting effective superpotential will turn out to
be those associated with the potential of certain elliptic
Calogero-Moser model \cite{Olshanetsky:am} with well defined
coefficients. The following table I shows the detailed
correspondence.

\subsection{$SL(2,Z)$ on Charge Lattice and Modular Transformation }

In the first subsection, we will briefly review how $SL(2,Z)$ acts
on $\CN=1^*$ theories on $S^1$ with untwisted boundary conditions.
Apart from reviewing known material and also reinforce the
inheritance of $SL(2,Z)$ by $\CN=1^*$ theories, this will also
identify the right modular transformation on the $\CN=1^*$ fields.
{}From the second subsections, we will turn to twisted cases.

The electromagnetic duality in the $\CN=4$ supersymmetric field
theory of Lie algebra $\CG$ on $R^{1+3}$ is simplest to see from
the electromagnetic charge spectrum of the theory when the gauge
symmetry is maximally broken by adjoint scalar fields. In fact,
the only solid evidence of $SL(2,Z)$ for $\CN=4$ theories comes
from checking that the actual spectrum of solitons does respect
this symmetry, which consists of searching for and determining
degeneracy of BPS states of monopoles and dyons. Checking the
quantum spectrum explicitly was done in some simpler settings
\cite{N=4}, and the result is consistent with the $SL(2,Z)$
symmetry.\footnote{The emergence of such spectra is not obvious at
all. For instance, pure $\CN=2$ theories share essentially the
same classical BPS spectra with $\CN=4$ theories
\cite{Lee:1998nv}, yet counting of quantum BPS states has shown
the $\CN=2$ spectra are almost never invariant under the $SL(2,Z)$
\cite{N=2}.}

For simplicity, let us consider the case with a single adjoint
scalar field breaking the gauge symmetry \cite{Dorey:1996jh}. For
any simple root $\balpha_a$, the complex charge lattice is given
by
\be p_a \balpha_a + \tau g_a \balpha_a^* \;  \label{clattice}
\,,\ee
where electric charge $p_a$ and magnetic charge $g_a$ are integers
defining a BPS state and its mass would be
\be \sqrt{\frac{4\pi}{{\rm Im}\;\tau} }\; \bigl| (p_a \balpha_a
+\tau g_a \balpha_a^* )\cdot \bv \bigr| \ \label{cmass}\,,\ee
where $\bv$ is the expectation value of the Higgs field whose
kinetic energy is in the standard form without $1/e^2$ factor.
Here the magnetic charge $g_a$ is an integer as the soliton
solution has integer topology. The electric charge $p_a$ is an
integer due to the quantization.

The above charge lattice and mass for the theories with simply
laced groups like $A_r, D_r, E_6, E_7, E_8$ are invariant under
the $SL(2,Z)$ transformation of the coupling (\ref{sl2z1}) and the
charge lattice
\be \left( \begin{array}{c} g_a \\ p_a \end{array}\right)
\rightarrow \left(
\begin{array}{cc} a & c \\ b& d \end{array}\right)^{-1}
\left( \begin{array}{c} g_a \\ p_a \end{array}\right)\,, \ee
where $a,b,c,d\in Z$ such that $ad-bc=1$.

The charge lattice for a non simply laced algebra is invariant
under only a subgroup of the $SL(2,Z)$. As shown in
Ref.\cite{Dorey:1996jh}, the symmetry of the charge lattice for
the theories of $B_r, C_r, F_4$ is $\Gamma_0(2)$, and that for the
$G_2$ theory is $\Gamma_0(3)$, where $\Gamma_0(n)$ is the
subalgebra such that $c=0 \; {\rm mod} \; n$. This can be checked
easily by considering the generators of $\Gamma_0(n)$. The reduced
modular symmetry $\Gamma_0(2)$ for the $\CN=4$ theories with gauge
group $B_r, C_r$ is identical to that obtained from the M-theory
picture in the previous section. One may expect this symmetry is
exact for the $\CN=4$ supersymmetric Yang-Mills theories with non
simply laced algebra, which would imply many nontrivial
predictions on magnetic monopole spectrum.

We note that there is an interesting aspect of this $\Gamma_0(2)$
symmetry. For the theories with $B_r, C_r, F_4$ gauge group, there
are roots of two different lengths and the two length differs by
$\sqrt2$. Take the example of $B_r$ where,
\begin{equation}
\balpha_s^2=1,\qquad \balpha_l^2=2\,,
\end{equation}
so that
\begin{equation}
\balpha^*_s=2\balpha_s\,, \qquad \balpha^*_l=\balpha_l\,.
\end{equation}
If we shift $\tau$ by one half,
\be \tau \rightarrow \tau + \frac{1}{2}\,, \ee
and then the above charge lattice get shifted
\be (p_s+ g_s)\,\balpha_s+ ( 2\tau ) g_s \,\balpha_s\,, \qquad
\left(p_l +\frac{g_l}{2}\right)\,\balpha_l + (2\tau)
\frac{g_l}{2}\,\balpha_l \,,\ee
with mutually co-prime integer pairs $(p_s,g_s)$ and $(p_l,g_l)$.
This shifted charge lattice is easily seen to be invariant under a
S-duality transformation
\be 2\tau \rightarrow -\frac{1}{2\tau}\,, \ee
which interchanges between integer (half-integer) electric and
integer (half-integer) magnetic charge of short (long) roots in
$2\tau$ unit. The superpotential for the corresponding $\CN=1^*$
theory on $R^{1+2}\times S^1$ with periodic boundary condition has
modular weight two once we also transform
\be \bX \rightarrow \frac{\bX}{2\tau}\quad. \ee
This self-duality is not new. One can show that the sequence of
the transformations $\tau \rightarrow \tau+1/2$, $2\tau
\rightarrow -1/2\tau$ and $\tau \rightarrow \tau -1/2$ belongs to
the group $\Gamma_0(2)$ because of the relation
\be \left(\begin{array}{cc} 1 & \frac{1}{2} \\
0 & 1 \end{array} \right) \left(\begin{array}{cc} 0 & -\frac{1}{2} \\
2 & 0 \end{array} \right) \left(\begin{array}{cc} 1 & -\frac{1}{2}
\\ 0 & 1
\end{array} \right)=\left(\begin{array}{cc} 1 & -1 \\
2 & -1 \end{array} \right) = ST^{-1}STS^{-1}\,.
\label{bcgamma2}\ee
Here $S$ and $T$ are some $SL(2,Z)$ generators such that $a=d=0$
and $-b=c=1$ for $S$, and $a=b=d=1$ and $c=0$ for $T$. A similar
construction holds for $C_r$ and $F_4$. For $C_r$, this is
precisely the ``$S$" transformation that leaves $\widetilde{O3}^+$
orientifold invariant. The theory with $G_2$ gauge group has a
similar transformation in $\Gamma_0(3)$.

In addition, the $S$-duality between the $\CN=4$ theories of $B_r$
and $C_r$ gauge groups would be clear when one uses the
normalization for the untwisted $C_r$ case such that its long
roots have length square 4 and its short roots have length square
2. This normalization fits well with the string theory point of
view where the simple roots of length square two originate from
the fundamental strings connecting D branes and the simple roots
of different length come from the orientifolds at the boundary.

Now the $\CN=2^*$ and $\CN=1^*$ theories for non simply laced
gauge groups should inherit this modular transformation properties
under the subgroup of $SL(2,Z)$. Indeed the reduced modular
symmetry for the integrable models associated with the $\CN=2^*$
theories, noticed in Ref.\cite{D'Hoker:1998yi}, turns out to be
identical to that of the charge lattice noticed above. Here, the
Weierstrass function $\wp(z)=\wp(z;2\pi i, 2\pi i \tau)$
transforms as a modular function of weight two as shown in
Appendix F. Depending on the Lie algebra $\CG$, we need to
consider a subgroup of $SL(2,Z)$ with accompanying the superfield
transformation,
\be \bX \rightarrow \frac{\bX}{c\tau+d} \,.\ee

When we compactify one of spatial directions, the charged
particles have logarithmically divergent energy and so the charge
lattice has no direct physical meaning, even though the $SL(2,Z)$
may remain meaningful. As the superpotential of the $\CN=1^*$
theories with periodic boundary condition is characterized by the
same integrable model as that of the corresponding $\CN=2^*$
theories, the superpotential would have the same reduced modular
properties. Indeed it is straightforward to see that the
superpotential (\ref{untpot}) for the untwisted theories with
simply laced groups $A_r,D_r,E_6,E_7,E_8$ has the $SL(2,Z)$
modular symmetry. The superpotential (\ref{untpot}) for the
untwisted theories with gauge groups $B_r,C_r,F_4$ has the
$\Gamma_0(2)$ modular symmetry. Finally the superpotential
(\ref{untpot}) for the untwised theory with $G_2$ gauge group has
the $\Gamma_0(3)$ modular symmetry.

In addition the $S$-duality between the $\CN=1^*$ theories of
$B_r$ and $C_r$ gauge groups can be seen clearly in the string
normalization. To see this in the effective superpotential
(\ref{untpot}), we rescale the parameters for $C_r$ case so that
\be \sqrt{2}\balpha\,, \sqrt{2}\bX\,, 2\tau\,, \CW
\quad\rightarrow \quad\balpha\,, \bX\,, \tau\,, \CW/2\,.\ee
The newly normalized superpotential for the untwisted $C_r$ case
becomes
\be \CW_{C_r^{(1)}}^{(nn)} = \sum_{\balpha \in {\cal
R}^+(\balpha^2=2)} \wp(\balpha\cdot\bX; 2\pi i, 2\pi i\tau )+
\sum_{\balpha \in {\cal R}^+(\balpha^2=4)}
\frac{1}{2}\wp(\frac{1}{2}\balpha\cdot\bX; 2 \pi i , \pi i\tau
)\,, \ee
which is S-dual to the superpotential for the untwisted $B_r$
theory,
\be \CW_{B_r^{(1)}} = \sum_{\balpha \in {\cal R}^+(\balpha^2=2)}
\wp(\balpha\cdot\bX; 2\pi i, 2\pi i\tau )+ \sum_{\balpha \in {\cal
R}^+(\balpha^2=1)} 2\wp(2\balpha\cdot\bX; 2 \pi i , 4\pi i\tau
)\,, \ee
as $\CW^{nn}_{C_r^{(1)}}(\bX/\tau) = \tau^2 \CW_{B_r^{(1)}}(\bX)$.
Here we dropped the mass parameter $m^3$ for simplicity.

\subsection{$ A_{2r}^{(2)}$ Case }

For $A_{2r}^{(2)}$ affine algebra, the subalgebra
$\CG_0=B_r=SO(2r+1)$ of even generators of $A_{2r}$ have $r^2$
positive roots,
\be {\rm medium \;\; roots} \;\;\; \bbeta_m= \frac{\bbe_a\pm
\bbe_b}{\sqrt{2}} \;\; (a<b), \;\; {\rm short\;\; roots} \;\;\;
\bbeta_{s}=\frac{\bbe_a}{\sqrt{2}} \,,\ee
with $\bbe_a$ ($a=1,...,r$) being orthonormal vectors. The subset
$\CG_1$ of odd generators of $A_{2r}^{(2)}$ belongs to the
symmetric traceless representation of $\CG_0$ and has $r^2+r$
positive root vectors,
\be {\rm long\;\; roots} \;\;\; \bbeta_l=\sqrt{2}\bbe_a, \;\;\;
{\rm medium\;\; roots } \;\;\; \bbeta_m, \;\; {\rm short\;\;
roots} \;\;\; \bbeta_s\,. \ee
Clearly the root vectors of $\CG_1$ is that of the so-called $BC$
system, which are made of the roots of $B_r$ and $C_r$ algebras.

Let us now express the superpotential (\ref{monopot}) for
$A_{2r}^{(2)}$ case. The contribution from the medium roots of
even and odd parts is
\be 2 \sum_{\bbeta_m} \left\{ \wp(2\bbeta_m\cdot \bX ; 2\pi i ,
4\pi i \tau) + \wp(2\bbeta_m\cdot \bX+ 2\pi \tau; 2\pi i , 4\pi i
\tau) \right\} = 2\sum_{\bbeta_m} \wp(2\bbeta_m\cdot \bX )\,, \ee
by the half period formula up to a $ \tau$-only function. {}From
now on the Weierstrass elliptic function $\wp(z;2\pi i , 2\pi
i\tau)$ is denoted as $\wp(z)$ for simplicity. Similarly, the
contributions from the short roots of even and odd parts become
\bear && 4\sum_{\bbeta_s} \left\{ \wp(4\bbeta_s\cdot \bX; 2\pi i ,
8\pi i \tau) +\wp(4\bbeta_s \cdot \bX+4\pi i \tau; 2\pi i ,8\pi i
\tau) \right\} \nonumber \\
&& \;\;\;\; = \sum_{\bbeta_s} \wp(2\bbeta_s \cdot \bX; \pi i, 2\pi
i \tau) = \sum_{\bbeta_s} \left( \wp(2\bbeta_s\cdot \bX) +
\wp(2\bbeta_s\cdot \bX+ \pi i ) \right) \label{a2rss} \,,\eear
by scaling and half-period formula up to a $\tau$-only function.
The contribution from the long roots, which only comes from odd
generators, is
\be \sum_{\bbeta_l} \wp(\bbeta_l \cdot \bX+\pi i \tau; 2\pi i ,
2\pi i \tau) \label{a2rlo}\,.\ee
The potential with the general coefficients for long, short and
shortest roots have appeared in
Ref.~\cite{Inozemtsev,Bordner:1998sw} as the twisted elliptic
Calogero model for the $BC$ system.

Now putting all contributions together and noticing $2\bbeta_s=
\bbeta_l$, we obtain the following expression for the
superpotential,
\be {\cal W}_{A_{2r}^{(2)}} = \sum_{\bbeta_m} 2 \wp(2\bbeta_m\cdot
\bX) + \sum_{\bbeta_l} \biggl( \wp(\bbeta_l\cdot \bX) +
\wp(\bbeta_l\cdot \bX+ \pi i ) + \wp(\bbeta_l\cdot \bX+\pi i \tau)
\biggr) \label{a2r}\,,\ee
where the sum is for all positive medium and long roots. Note that
the funny looking range (\ref{ranget1}) and (\ref{zrange}) of the
bosonic part $\bz$ of $\bX$ in the $A_{2r}^{(2)}$ theory is just
suitable for the above expression for the superpotential as it is
periodic under $ \bbeta_l\cdot\bz \sim \bbeta_l\cdot \bz + \pi i
\sim \bbeta_l\cdot\bz+ \pi i \tau$. The Weyl reflection is
manifest.

The $SL(2,Z)$ symmetry of the above potential can be seen as
follows: the first two sums each have the right modular properties
under the $SL(2,Z)$. For the rest, note that $S$, which maps
$\tau\rightarrow -1/\tau, \bX\rightarrow \bX/\tau$, exchanges the
last two terms up to the modular transformation of weight two. On
the other hand, $T$, which transforms $\tau\rightarrow \tau+1$,
$\bX \rightarrow \bX + \pi i \sum_{a=1}^{r} \bbe_a /\sqrt{2}$,
also interchanges the second and third expressions. Since $S$ and
$T$ generate $SL(2,Z)$, the sum of the last three terms together
have the right modular properties.

\subsection{$ A_{2r-1}^{(2)}$ Case}

For the $A_{2r-1}^{(2)}$ case, the $\CG_0=C_r= Sp(r)$ of even
generators has $r^2$ positive generators,
\be {\rm short\;\; roots}\;\;\; \bbeta_s=\frac{\bbe_a\pm
\bbe_b}{\sqrt{2}} \;\;\; (a<b) , \;\;\;\;\; {\rm long \;\; roots}
\;\;\; \bbeta_l=\sqrt{2} \bbe_a\; , \ee
where $\bbe_a$ with $a=1,...,r$ are orthonormal vectors. The
$\CG_1$ of odd generators belongs to the $2r^2+r$ dimensional
representation and has $r^2-r$ positive root vectors,
\be {\rm short\;\; roots}\;\;\;\bbeta_s \; .\ee

The contribution to the superpotential (\ref{monopot}) from the
monopole instantons associated with the long roots only in the
$\CG_0$ is
\be \sum_{\bbeta_l} \wp( \bbeta_l \cdot \bX; 2\pi i , 2\pi i
\tau)\,. \ee
The contribution to the superpotential from the monopole
instantons associated with the short roots of the even and odd
generators is
\bear & & 2\sum_{\bbeta_s} \biggl( \wp (2\bbeta_s \cdot \bX; 2\pi
i , 4\pi i \tau) +\wp (2\bbeta_s \cdot x + 2\pi
i \tau ; 2\pi i , 4\pi i \tau) \biggr) \nonumber \\
& & \;\; = 2\sum_{\bbeta_s} \wp(2\bbeta_s \cdot \bX; 2\pi i , 2\pi
i \tau)\,, \eear
up to a $\tau$ dependent function after using the half-period
formula. Thus, the holomorphic superpotential is
\be {\cal W}_{A_{2r-1}^{(2)}} = 2\sum_{\bbeta_s} \wp(2\bbeta_s
\cdot \bX) + \sum_{\bbeta_l} \wp( \bbeta_l\cdot \bX)
\label{a2rm1}\,,\ee
where the sum is only for the positive roots. The $SL(2,Z)$
property is manifest. In addition we note that it is associated
with untwisted elliptic Calogero-Moser model for the Lie algebra
$B_r=SO(2r+1)$, whose extended Dynkin diagram is the diagram for
the simple coroots of $A_{2r-1}^{(2)}$.

\subsection{$ D_{r+1}^{(2)} $ Case }

For the $D_{r+1}^{(2)}$, the $\CG_0=B_r$ of even generators has
$r^2$ positive roots
\be {\rm long\;\; roots} \;\;\; \bbeta_l=\bbe_a\pm \bbe_b \;\;
(a<b) ,\;\; {\rm short\;\; roots }\;\;\; \bbeta_s= \bbe_a\,, \ee
where $\bbe_a,a=1,2,...,r$ are $r$ orthonormal vectors. The
$\CG_1$ of odd generators belongs the $2r+1$ dimensional vector
representation and has $r$ positive roots
\be {\rm short\;\; roots}\;\;\; \bbeta_s \,.\ee
The contribution to the superpotential (\ref{monopot}) from even
long generators of $\CG_0$ is
\be \sum_{\bbeta_l} \wp(\bbeta_l \cdot x; 2\pi i , 2\pi i \tau)\,.
\ee
The contributions from the even and odd short generators are
\be 2\sum_{\bbeta_s} \left\{ \wp(2\bbeta_s \cdot \bX; 2\pi i ,
4\pi i \tau) + \wp (2\bbeta_s\cdot \bX+ 2\pi i \tau; 2\pi i , 4\pi
i \tau) \right\} = 2\sum_{\bbeta_s} \wp(2\bbeta_s\cdot \bX)\,, \ee
by the half-period formula up to the $\tau$ only function.
Together the superpotential for the $D_{r+1}^{(2)} $ case is
\be {\cal W}_{D_{r+1}^{(2)}}= \sum_{\bbeta_l} \wp(\bbeta_l \cdot
\bX)+ 2\sum_{\bbeta_s} \wp(2\bbeta_s\cdot \bX) \label{d2r}\,, \ee
which is manifestly $SL(2,Z)$ invariant modulo weight two. This
potential is obviously associated with the untwisted elliptic
Calogero-Moser system for the Lie algebra $C_r$ whose extended
Dynkin diagram is identical to that of the coroots of
$D_{r+1}^{(2)}$.

\subsection{$E_6^{(2)}$ and $D_4^{(3)}$ Cases}

Let us consider the $E_6^{(2)}$ case first. For the $E_6^{(2)}$ of
dimension 78, the Lie algebra of even generators are identical to
$\CG_0= F_4$ of dimension 52. The odd part $\CG_1$ is of 26
dimensional representation of $F_4$. In terms of four orthonormal
vectors $\bbe_a$ with $a=1,2,3,4$ the positive root vectors for
even part $\CG_0=F_4$ of $E_6 $ are given as
\be {\rm long\; roots} \; \bbeta_l=\bbe_a\pm \bbe_b \; (a<b), \;\;
{\rm short\; roots} \;\; \bbeta_s= \bbe_a , \; {\rm or} \;
\frac{1}{2}(\bbe_1\pm \bbe_2\pm \bbe_3\pm \bbe_4)\,. \ee
The positive root vectors for odd part $\CG_1$ of $E_6^{(2)}$ are
given as
\be {\rm short\;\; roots} \;\; \bbeta_s \,.\ee
Thus the even long root contribution to the superpotential
(\ref{monopot}) is
\be \sum_{\bbeta_l} \wp(\bbeta_l\cdot \bX; 2\pi i , 2\pi i
\tau)\,. \ee
The sum of even and odd short root contributions is
\be 2 \sum_{\bbeta_s} \left\{ \wp(2\bbeta_s\cdot \bX; 2\pi i ,
4\pi i \tau)+ \wp(2\bbeta_s \cdot \bX+2\pi i \tau; 2\pi i , 4\pi i
\tau) \right\}= 2 \sum_{\bbeta_s} \wp(2\bbeta_s \cdot \bX)\,, \ee
again up to a $\tau$-only function by the half-period formula.
Putting them together we have the effective superpotential
\be {\cal W}_{E_6^{(2)}} = \sum_{\bbeta_l} \wp(\bbeta_l \cdot \bX)
+ 2 \sum_{\beta_s} \wp(2\bbeta_s \cdot \bX) \label{e62}\,,\ee
which has the manifest $SL(2,Z)$ property. In addition this
potential is associated with the untwisted elliptic Calogero-Moser
potential for the Lie group $F_4$, whose extended Dynkin diagram
is dual to that of $E_6^{(2)}$.


For $D_4^{(3)}$ of dimension 28, the $\CG_0=G_2$ of even
generators has $6$ positive roots,
\bear && {\rm long\;\; roots} \;\;\;\bbeta_l= \sqrt{2}\;\bbe_1,
\pm \sqrt{\frac{1}{2}} \;\bbe_1+ \sqrt{\frac{3}{2}}\;\bbe_2\,,
\nonumber
\\ && {\rm short \;\; roots }\;\; \bbeta_s = \sqrt{\frac{2}{3}}\;\bbe_2, \pm
\sqrt{\frac{1}{2}}\; \bbe_1+\sqrt{\frac{1}{6}} \;\bbe_2 \,,\eear
The $\CG_1$ and $\CG_2$ have each three short positive roots,
which are identical to the short roots of $\CG_0$.

The contribution to the superpotential (\ref{monopot}) from the
even long roots is
\be \sum_{\bbeta_l} \wp ( \bbeta_{l}\cdot \bX; 2\pi i, 2\pi
i\tau)\,, \ee
which is modular. The contribution from the even and twisted short
roots of $\CG_n$ with $n=0,1,2$ is
\be 3\sum_{\bbeta_s}\sum_{n=0}^2\left\{ \wp(3\bbeta_s\cdot \bX+
2\pi i \tau n; 2\pi i , 6\pi i \tau) \right\} = 3\sum_s \wp(
3\bbeta_s\cdot\bX; 2\pi i , 2\pi i \tau)\,, \ee
up to a $\tau$ only function due to the triple period formula. The
total superpotential is then
\be {\cal W}_{D_4^{(3)}} = \sum_{\bbeta_l} \wp ( \bbeta_{l}\cdot
\bX) +3\sum_{\bbeta_s} \wp( 3\bbeta_s\cdot\bX ) \label{d34} \,,\ee
where the sum is for all positive long and short roots of $\CG_n$
of $D_4^{(3)}$. The result is again modular with weight two. It is
also associated with the untwisted Calogero-Moser potential for
the Lie algebra $G_2$ whose extended Dynkin diagram is dual to
that of $D_4^{(3)}$.

\section{Glueball Superpotential }

We have found the $SL(2,Z)$ modular superpotentials for all
$\CN=1^*$ theory with twisted boundary condition. The
superpotentials we wrote are functions of the chiral superfields,
${\bf X}$, which are essentially the photon supermultiplet that
remains unbroken by the Wilson line symmetry breaking. In turn
this superpotential should lead to the gluino condensation,
determining the vacuum structure and its accompanying value of the
gluino condensate, among many other physical quantities.

Let us first recall how this happens. The superpotential is
expected to have serveral discrete stationary points $\{ \bX_0 \}$
in the fundamental domain, where
\be \frac{\partial { \cal W} }{\partial X_a}= 0 \,, \;\;
a=1,..,r'\,. \ee
The number of supersymmetric ground states is the number of such
stationary points modulo the Weyl reflections. The values of the
superpotential at these stationary points would be holomorphic
functions of only $m$ and $\tau$ variables. The derivative of the
superpotential with respect to $\tau$ would lead to the
expectation value of the glueball superfield $S$,
\be \langle S \rangle = \frac{1}{2\pi i } \frac{\partial {\cal
W}(\bX_0)}{\partial \tau}\,, \ee
at a given vacuum $\bX=\bX_0$. This relation can be seen from the
fact that the bare Lagrangian contains the term
\begin{equation}
2\pi i \tau S \bigr\vert_{\,\theta^2} +\hbox{complex conjugate}\,,
\end{equation}
which is nothing but the kinetic term for the gauge field.

This can be made more precise as follows. We will show how this
goes for our (twisted) $N=1^*$ theories, but similar constructions
are possible in other cases also. With the $\CG_0$ of the rank
$r'$, the twisted affine algebra $\CG^{(L)}$ has $r'+1$
fundamental monopoles corresponding to the simple roots $\bbeta_a,
a=0,1,2...r'$ of the affine algebra. We define $r'$ independent
chiral superfields
\be Y_a= \bbeta_a^*\cdot \bX \,,\ee
plus one more variable
\be Y_0= \frac{4\pi i \tau }{L\bbeta_0^2} + \bbeta^*_0\cdot
\bX\,.\ee
Due to the properties of the simple roots discussed in Appendix E,
we have the constraint
\be \sum_{a=0}^{r'} \tilde{k}_a^* Y_a=\frac{4\pi i \tau
\tilde{k}_0^*}{L\bbeta_0^2}\,, \ee
with comarks $\tilde{k}_a^*$ for $\CG^{(L)}$, which, according to
(\ref{fact2}), is
\be \sum_{a=0}^{r'} \tilde{k}_a^*Y_a=\frac{4\pi i \tau
}{\balpha_0^2} =2\pi i \tau\,, \ee
as the lowest root $\balpha_0$ of $\CG$ has the length square two
in our convention. In terms of their exponents,
\be y_a = e^{Y_a} ,\; a=0,1,...r'\,,\ee
the above constraint becomes
\be \prod_{a=0}^{r'} y_a^{\tilde{k}_a^*} = q \label{insmono}\,,
\ee
with $q=e^{2\pi i \tau}$ as discussed in subsection (3.3).

In small coupling constant $e^2$ limit, which is equivalent to
large $-{\rm Re}(2\pi i \tau)$, or small $|q|$ limit, we are
expanding the superpotential with a chiral superfield lying near
the stationary points of the superpotential (\ref{etoda}), which
we will see later that
\be \bX \approx 2\pi i \tau \frac{\brho}{h} + \delta \bX
\label{stationary0} \,,\ee
with order one quantity $\delta \bX$. $\brho$ is the Weyl vector
of the twisted algebra
\be \brho = \sum_{a=1}^{r'} \bw_a \,,\ee
with the fundamental weights $\bw_a$ such that
$\bw_a\cdot\bbeta_b^* = \delta_{ab}$. Around this background, we
see that
\be y_a \sim q^{1/h} , a=0,1,2,...,r'\,,\ee
where for the estimation of $y_0$ we have used the properties of
the roots as discussed in Appendices D and E.

Now, we are interested in expressing our potential in series as a
sum over all possible magnetic monopole instantons and pure
instanton corrections. One can make a general statement about all
twisted affine algebra with a minor modification for
$A_{2r}^{(2)}$ case. In all other cases where $\beta_0$ is a short
root with $\tilde{k}_0^*=1$, a positive co-root $\bbeta^*$ in any
level $\CG_n$ can be expressible as a nonnegative integer sum of
the simple co-root $\bbeta_a^*$ of $\CG_0$,
\be \bbeta^*= \sum_{a=1}^{r'} n_a \bbeta_a^* \,\,, \;\; n_a\;\;
{\rm non \; negative \; integer} \label{nnumbers} \,,\ee
and $ \sum_a^{r'} n_a \le h-1$, where the inequality is saturated
when (\ref{nnumbers}) becomes $-\bbeta_0^*$. The superpotential is
a sum of Weierstrass functions whose arguments are $\bbeta^*\cdot
\bX$. Take any one of these functions and expand it in small $q$,
and we find some general series like
\be \wp(\bbeta^*\cdot \bX) = \frac{1}{12}+ \sum_{k=0}^\infty ky^k
+\sum_{k=1}^\infty \sum_{n=1}^\infty kq^{kn}(y^k+y^{-k} -2) \,,\ee
where $y=e^{\bbeta^*\cdot\bX} $. From the $\bbeta^*$ decomposition
(\ref{nnumbers}) in terms of simple roots, we also have
\be y = \prod_{a=1}^{r'} y_a^{n_a} \,.\ee
As we have $q$ expressed in terms of $y_a$ as in
Eq.(\ref{insmono}), and noting that $qy^{-1}$ can always be
expressed in positive powers in $y_a$'s, we can see the
superpotential as an infinite sum over multi-monopole-instanton
contributions
\be {\cal W} = m^3 \sum_{K=\{n_a\}} c_K \prod_{a=0}^{r'}
y_a^{n_a}\,, \ee
where the sum is for all possible sets $K=\{n_a\}$ of non negative
integers $n_a$ and the coefficients $c_K$ are determined by our
$SL(2,Z)$ invariant superpotential. For the $A_{2r}^{(2)} $ case,
whose Weierstrass function involves shifted arguments, we will
also see later that all corrections can be expressed as the
products of the contributions from the fundamental monopoles.

This expression is not entirely correct since not all of $Y_a$'s
are dynamical variables. In order to have a sensible
superpotential we must introduce a Lagrange multiplier field $S$
as follows \be {\cal W}(Y_a, S;\tau) = m^3 \sum_{K=\{n_a\}} c_K
\prod_{a=0}^{r'} y_a^{n_a} - S(-2\pi i \tau + \sum_{a=0}^{r'}
\tilde{k}_a^* Y_a ) \label{gbseries1}\,, \ee
which removes one variable $\sum_{a=0}^{r'} \tilde{k}_a^* Y_a $
and replace it by one parameter $\tau$. It is now quite obvious
that $S$ is nothing but the glueball superfield, and the gluino
condensate is given by \be \langle S \rangle \sim \frac{\partial
{\cal W}}{\partial \tau}\,, \ee
from the equation of motion for $S$. It is then a matter of
integrating out the superfields $Y_a$ to obtain the glueball
superpotential. \be \CW(S;\tau)={\cal W}(Y_a,
S;\tau)\biggr\vert_{\hbox{ extremize w.r.t. $Y_a$'s}}\,. \ee This
sort of duality or mirror symmetry between two types of
superpotential has appeared first in Ref.\cite{Hori:2000kt} and
has been used extensively in a recent work \cite{Aganagic:2003xq}.

In this section, we will show that this glueball superpotential
for $\CN=1^*$ theory with twisted boundary condition matches
exactly with that for the $\CN=1^*$ theory with periodic boundary
condition in leading terms in the series expansion. This might
sound a bit odd initially, given that we are dealing with two
theories of completely different gauge groups and matter content,
for example in the small radius limit. The first subsection will
motivate this identity and explain why this must be the case.

In particular, when the dual Coxeter number $h$ is small enough,
the pure instanton correction of order $q$ is included in this
comparison. The pure instanton correction for the twisted case is
fixed by the $SL(2,Z)$ symmetry, so this comparison of two
glueball superpotentials may be regarded as an independent check
of our prescription.

\subsection{$\CW^{VY}$ and Universality}

In the leading order (or to order $q^{\frac{1}{h}}$) we keep only
terms linear in $y_a$, or the corrections from the fundamental
monopoles. The superpotential (\ref{etoda}) in the $\CN=1$ theory
of the twisted affine algebra $\CG^{(L)}$ can be rewritten as
\be \CW_{N=1}(Y_a;S) = m^3\biggl( \sum_{a=0}^{r'}
\frac{2}{\bbeta_a^2}\; y_a \biggr)-S(-2\pi i \tau +
\sum_{a=0}^{r'} \tilde{k}_a^* Y_a) \label{etoda1} \,,\ee
where $\tilde{k}_a^*$ is the comarks of the twisted Lie algebra.
Integrating over the glueball field $S$ renders $Y_0$ to be a
linear combination of other $Y$'s, leading to the Toda potential
(\ref{etoda}).

Integrating over $Y_a$ instead leads to a glueball superpotential
\be {\cal W}(S;\tau) = hS -S \ln \left(
\frac{S}{\Lambda^3}\right)^h \,,\ee
where $h$ is the dual Coxeter number of $\CG$ and the confinement
scale in the Coulomb phase is now given as
\be \Lambda^3 = m^3 q^{\frac{1}{h}} \prod_{a=0}^{r'} \bigg[
\frac{\tilde{k}_a^*\bbeta_a^2}{2}\biggl]^{\tilde{k}_a^*}\,. \ee
The stationary point of the above superpotential leads to the
glueball field expectation value (\ref{glueball1}) in $h$ number
of vacua. For a given vacuum, the expectation value of $Y_a$ or
$y_a$ can be read directly from the above action (\ref{etoda1}),
which are
\be \langle Y_a \rangle = \frac{2\pi i k}{h} + \ln\left(
\frac{\tilde{k}_a^* \bbeta_a^2\Lambda^3}{2m^3} \right),\;
k=0,1,...,(h-1)\,, \ee
and so $\langle y_a \rangle \sim q^{1/h}$, which is consistent.
Indeed these stationary points are consistent with the assumption
for the stationary points (\ref{stationary0}) of the chiral
superfield $\bX$. The weak coupling series of the glueball
superpotential leads to the corrections to these stationary points
by a similar series in powers of $q^{1/h}$.

Note that the form of the superpotential of the twisted theory
$\CG^{(L)}$ is the same as that of the untwisted theory $\CG$,
namely the Veneziano-Yankielowicz superpotential. Furthermore, the
identities in Appendix E shows that the confinement scale of the
$\CN=1$ theory with twisted boundary condition is identical to
that of the $\CN=1$ theory with periodic boundary condition,
resulting in the identical glueball potential.

We cannot possibly say the same thing about the superpotential
$\CW(\bX)$. To begin with, the untwisted ${\CG}$ theory and the
twisted ${\CG}^{(L)}$ ($L>1$) theory have different ranks, and
therefore their superpotentials are analytic functions of
different number of complex variables. Thus, it must be quite
surprising to see that one and the same glueball superpotential is
found when we started with two such different superpotential.

A hint of why this happens can be found in the fact that
superpotential $\CW(\bX)$ has no explicit dependence on the
compactification radius. If we had not known better, this might
lead us to conclude that the superpotential is independent of the
twist, the argument being that a large radius limit can always
wash out any boundary condition. However, this is not the case
with $\CW(\bX)$ as we just explained.\footnote{This observation
seems at odd with the notion that at least in the larger radius
limit the physics must be independent of boundary condition. What
must be happening here is that $\CW(\bX)$ of the twisted theory is
the result of integrating out more degrees of freedom than
$\CW(\bX)$ of the untwisted theories, and in the large radius
limit, the former is less reliable as a low energy
superpotential.} $\CW(\bX)$ of the twisted theory do retain the
memory of the twisting in that it is defined with respect to
Cartan sector of $\CG_0$ instead of entire $\CG$.

The glueball superpotential, however, is defined as over a single
chiral field, $S$, whose definition involves tracing over all
gauge indices. This means that $\CW(S;\tau)$ does not retain the
memory of the twist in the way $\CW(\bX)$ does, and is really
insensitive to the boundary condition in the large radius limit.
Unlike $\CW(\bX)$, then, the glueball superpotential must be
insensitive to the twist at all radius, which explains why we find
the identical Veneziano-Yankielowicz superpotential $\CW^{VY}$ for
the twisted and the untwisted cases alike. We expect this exact
matching of glueball superpotentials of twisted and untwisted
theory to persist to all orders beyond $\CW^{VY}$. The rest of
this section is devoted to checking this to the 4th power in $S$.

\subsection{Glueball Superpotential of Untwisted $ADE$ Theories}

Let us compute the glueball superpotential of untwisted theories
with simply laced groups, $A_r$, $D_r$ and $E_6,E_7,E_8$ to order
$S^4$. As we noted already, all twisted theories arise from one of
these simply laced cases, and in the following subsections we will
compare the results here with their twisted counterpart.

It is convenient to introduce the extended Cartan matrix
$\hat{C}_{ab} $ and the symmetric incidence matrix $\CI_{ab}$ such
that $\hat{C}_{ab} = 2\balpha_a\cdot \balpha_b^* =2\delta_{ab}
-\CI_{ab}\,,$ where $a,b=0,...,r$. The incidence matrix $\CI_{ab}$
is zero (one) when two simple roots $\balpha_a$ and $\balpha_b$
are disconnected (connected) in the extended Dynkin diagram. The
diagonal elements of the incidence matrix vanish. The simple roots
are $\balpha_a$ with $a=0,1,2...r\,, $ where $r$ is the rank of
the algebra. The dual Coxeter number is $h$ and we define
$Y_0=2\pi i\tau -\balpha_0\cdot \bX$ and
$Y_a=\balpha_a\cdot\bX\,\,, a=1,2...r$.

The superpotential is
\be \CW = \sum_{\balpha>0} \wp(\balpha^*\cdot x)\,, \ee
where the sum is over the positive roots of the respective $ADE$
algebra. This may be expanded in a series of the contributions to
the fundamental monopole instantons with $y_a=e^{Y_a}$ as above.
Keeping terms up to order $q^{\frac{4}{h}}$, we find that the
series becomes\footnote{For the rest of this section, we drop the
mass parameter $m^3$ for simplicity.}
\bear \CW_{A^{(1)},D^{(1)},E^{(1)}} = && \sum_{a=0}^r ( y_a +
2y_a^2+3y_a^3+4y_a^4 ) + \frac{1}{2} \sum_{ab} \CI_{ab}y_ay_b +
\frac{1}{2}\sum_{abc}I_{ab}I_{bc}y_ay_by_c \nonumber \\
&&-\frac{1}{2}\sum_{ab}I_{ab}y_a^2 y_b +
\frac{1}{2}\sum_{abcd}I_{ab}I_{bc}I_{cd}y_ay_by_cy_d-
\sum_{abc}I_{ab}I_{ac}y_a^2y_by_c\nonumber \\
&&+\frac{3}{2}\sum_{ab}I_{ab}y_a^2y_b^2+
\frac{1}{6}\sum_{abcd}I_{ab}I_{ac}I_{ad}y_ay_by_cy_d -
\frac{1}{2}\sum_{abc}I_{ab}I_{ac}y_ay_b^2y_c\nonumber \\&&+
\frac{1}{3}\sum_{ab}I_{ab}y_ay_b^3 -S(-2\pi i \tau+ \sum_{a=0}^r
k_a^*Y_a )\,, \eear
where $k_a^*$ is the comarks of the untwisted affine algebra.
After integration over $Y_a$, we obtain the superpotential for the
glueball superpotential to order $S^4$,
\bear\CW_{A^{(1)},D^{(1)},E^{(1)}}(S;\tau)
 &=& \CW^{VY}_{A,D,E} \nonumber \\
&+&3h_2 S^2-14 h_3 S^3 + (115h_4 -
\frac{23}{2}\hat{C}_{ab}(k_a^*)^{2}(k_b^*)^{2}) S^4 \nonumber
\\
&+&{\cal O}(S^5) \label{DE3glue}\,, \eear
where $h_n = \sum_{a=0}^r (k_a^*)^n$ with the comarks $k_a^*$.

Note that for all simply laced groups, $\hat{C}_{ab}k_b^* =0 $.
For $A_r$, $k_a^*=1$ for all $a=0,1,...,r$, and so the last term
is absent for the theory with $A_r$. This last term proportional
to $C_{ab}(k_a^*)^2(k_b^*)^2$ is nontrivial for the other, $D$ and
$E$ type, simply laced algebra. This particular term appears
missing in the result of Ref.~\cite{Aganagic:2003xq}.

\subsection{$A_r^{(1)}$ vs. $A_r^{(2)}$}

Let us write out the case of $A_r$ series more explicitly here.
For untwisted $A_r$ theory, the simple roots are
\be \balpha_a = \bbe_a - \bbe_{a+1} , \; a=0,1,...,r\,, \ee
where $\bbe_{r+1}=\bbe_0$. With $Y_0=2\pi i \tau-\balpha_0\cdot
\bX$ and $Y_a = \balpha_a\cdot \bX, a=1,2,...r$, and $y_a=
e^{Y_a}$, we can expand the above superpotential in a series, to
order $q^{8/h}$. Writing out the above expansion of the
superpotential for $A_r$ untwisted theory, we find
\bear \CW_{A_r^{(1)}} = && \sum_{a=0}^r \left( y_a +
2y_a^2+3y_a^3+4y_a^4+5y_a^5+6y_a^6 + 7y_a^7+8y_a^8 +y_ay_{a+1} +
2y_a^2y_{a+1}^2 \right.
\nonumber \\
&& \;\;\; +3y_a^3y_{a+1}^3+ 4y_a^4y_{a+1}^4 + y_ay_{a+1}y_{a+2}+
2y_a^2y_{a+1}^2y_{a+2}^2+ y_ay_{a+1}y_{a+2}y_{a+3}
\nonumber \\
&& + 2y_a^2y_{a+1}^2y_{a+2}^2y_{a+3}^2 +
y_ay_{a+1}y_{a+2}y_{a+3}y_{a+4} +
y_ay_{a+1}y_{a+2}y_{a+3}y_{a+4}y_{a+5}
\nonumber \\
&& + y_ay_{a+1}y_{a+2}y_{a+3}y_{a+4}y_{a+5}y_{a+6}
y_ay_{a+1}y_{a+2}y_{a+3}y_{a+4}y_{a+5}y_{a+6}y_{a+7}
\nonumber \\
&& \left. +
y_ay_{a+1}y_{a+2}y_{a+3}y_{a+4}y_{a+5}y_{a+6}y_{a+7}y_{a+8}
\right) -S(-2\pi i \tau+ \sum_{a=0}^r Y_a ) \,.\eear
Here we put the periodic condition on the indices so that
$y_{r+1}=y_0$ and so on. Solving the equation for $Y_a$, we get
the glueball superpotential in a series,
\bear \CW_{A_r^{(1)}}(S;\tau) = && \CW^{VY}_{A_r} + h\bigl( 3S^2
-14
S^3 + 115 S^4 - 1206 S^5 \nonumber \\
&& + \frac{72576}{5} S^6 - 190968 S^7+ \frac{18721233}{7} S^8
\bigr)+ {\cal O}(S^9) \,.\eear
This particular series is valid up to $8$-th power in $S$ for the
cases with dual Coxeter number $h=r+1$ larger than $8$.

Now let us turn to the twisted cases. First, take the case of
$A_{r=2r'}^{(2)}$, whose zero degree subgroup is $\CG_0 =
Sp(r')=USp(2r')$ of rank $r'$ and the dual Coxeter number is
$h=2r'+1=r+1$. For $A_{2r'}^{(2)}$, the root vectors given in
orthonomal vectors were studied in the subsection (4.2). The
simple roots, dropping the degree, are
\be \bbeta_0=-\sqrt{2}\bbe_1\,,\;\; \bbeta_a= \frac{\bbe_a
-\bbe_{a+1}}{\sqrt{2}}\,, \;\; a=1,2,...(r'-1)\, ,\;\;
\bbeta_{r'}= \frac{\bbe_{r'}}{\sqrt{2}} \,,\ee
with $\bbe_a$ with $a=1,...,r'$ being orthonormal vectors.

The superpotential (\ref{a2r}) can be expanded in small $q$
series. With $Y_0 = \pi i \tau +\bbeta_0\cdot \bX$, $Y_a= \bbeta_a
\cdot \tau$ with $a=1,2,...r$, all variables $y_a=e^{Y_a} \sim
q^{\frac{1}{h}}$. For a medium size root $\bbeta_m$, we get the
series expansion. For the long positive root $\bbeta_l$
contribution $\wp(\bbeta_l\cdot \bX-\pi i \tau)$, we notice that
\be \pi i \tau -\bbeta_l\cdot \bX = Y_0 + \sum_{a=1}^{r'}n_a
Y_a\,, \ee
where $n_a$ are non negative integers. The reason is that a long
negative root $-\bbeta_l$ can be expressed as a sum of the lowest
long negative root $\bbeta_0$ plus simple roots. Thus the series
will be an infinite series of the contributions from the
fundamental monopoles. For the rest two contributions for a long
positive root, we employ the identity
\bear & & \wp(\bbeta_l\cdot \bX) + \wp(\bbeta_l\cdot\bX + \pi i )
\nonumber \\
& & \;\; = \frac{2}{12} \sum_{k=1}^\infty (1+ (-1)^k)
ke^{\bbeta_l\cdot \bX} + \sum_{k,m=1}^\infty k\;q^{km} \left( (1+
(-1)^k)(e^{\bbeta_l\cdot\bX} + e^{-\bbeta_l\cdot\bX}) -4\right)
\nonumber \\
&& =\;\; \frac{2}{12} + 4 \sum_k ke^{\bbeta_s^*\cdot\bX}+
4\sum_{k,m=1}^\infty kq^{2km} (e^{\bbeta_s^*\cdot \bX k} +
e^{-\bbeta_s^*\cdot\bX}) -4\sum_{k,m=1}^\infty k q^{km}\,,\eear
where $\bbeta_s^*=2\bbeta_l$. Since $q$ is a product of $y_a$'s
and $q^2e^{-\bbeta_s^*\cdot\bX}$ can be expressed in the positive
power products of $y_a$, the above contributions can be expressed
in a series of contributions from the fundamental monopoles. The
factor four is exactly the needed coefficient $2/\bbeta_s^2$.

Once that is done, we can find the superpotential in the series
expansion to order $q^{\frac{4}{h}}$,
\bear \CW_{A_{2r'}^{(2)}} &=& ( y_0 +2y_0^2+3y_0^3+4y_0^4) + 2
\sum_{a=1}^{r'-1} (y_a+2y_a^2+3y_a^3+4y_a^4) \nonumber \\
&& + 4(y_{r'}+2y_{r'}^2+3y_{r'}^3+4y_{r'}^4) +
(y_0y_1+2y_0^2y_1^2) + 2\sum_{a=1}^{r'-1} (y_ay_{a+1} +
2y_a^2y_{a+1}^2)\nonumber \\
&& + 4y_{r'-1}^2 y_{r'} + y_0y_1y_2 + 2y_0^2y_1
+2\sum_{a=1}^{r'-2}
y_ay_{a+1}y_{a+2} \nonumber \\
& & +y_0y_1y_2y_3 + 2y_0^2y_1y_2 + 2y_{r'-1}y_{r'-1}^2y_{r'}+
2\sum_{a=1}^{r'-3}y_ay_{a+1}y_{a+2}y_{a+3}
\nonumber \\
& & -S(-2\pi i \tau+ 2Y_0+2Y_1+\cdots + 2Y_{r'-1}+Y_{r'})\,. \eear
After integrating over $Y_a$ order by order in perturbation, we
obtain the glueball superpotential
\be \CW_{A_{2r'}^{(2)}}(S;\tau) = \CW^{VY}_{A_{2r' }} + h( 3S^2
-14 S^3 +115 S^4) + {\cal O}(S^5) \,.\ee
To the order $S^4$, this series matches exactly with the series
expansion obtained for the theory with untwisted affine
$A_{r=2r'}^{(1)} $ algebra above. The contributions from the
constant coefficient $1/12$ matches exactly between the twisted
and untwisted glueball superpotential due to the fact (E.6) of
Appendix E. Such match of constants occurs also for other twisted
cases due to the fact (E.7).

The second case to consider is $A_{r=2r'-1}^{(2)}$, whose zero
grading is $\CG_0=C_{r'}$ with rank $r'$ and the dual Coxeter
number $h=2r'=r+1$. The roots of this twisted affine algebra is
studied in Sec. (4.3). The simple roots are
\be \bbeta_0 = -\frac{\bbe_1+\bbe_2}{\sqrt{2}}\,, \;\; \bbeta_a=
\frac{\bbe_a-\bbe_{a+1}}{\sqrt{2}} \,,\;\; a=1,2...(r'-1)\,, \;\;
\bbeta_{r'} = \sqrt{2}\bbe_{r' }\,.\ee
With $Y_0= 2\pi \tau -\bbeta_0^*\cdot \bX$ and
$Y_a=\bbeta_a^*\cdot \bX, a=1,2,...r$, we introduce again the
contributions from the fundamental monopoles, $y_a= e^{Y_a} \sim
q^{\frac{1}{h}}$. The contributions from the short and long roots
to the superpotential (\ref{a2rm1}) can be now expressed in the
series of $y_a$. Expanding the superpotential (\ref{a2rm1}) to
order $q^{\frac{4}{h}}$, we obtain the series,
\bear \CW_{A_{2r'-1}^{(2)} }&=&2\sum_{a=0}^{r'-1} (y_a + 2y_a^2 +
3y_a^3+ 4y_a^4) + (y_{r'} + 2y_{r'}^2+3y_3^3+4y_{r'}^4) + 2(y_0y_2
+ 2y_0^2y_2^2) \nonumber \\ & & + 2 \sum_{a=1}^{r'-2} (y_ay_{a+1}
+ 2y_a^2y_{a+1}^2) + (y_{r'-1}y_{r'} + 2y_{r'-1}^2 y_{r'}^2)+
2y_0y_2y_3 \nonumber
\\
& & + 2\sum_{a=0}^{r'-3}(y_a
y_{a+1}y_{a+2})+ y_{r'-2}y_{r'-1}y_{r'}+2y_{r'-1}y_{r'}^2 + 2 y_0y_2y_3y_4 \nonumber \\
& & + 2\sum_{a=0}^{r'-4}(y_a
y_{a+1}y_{a+2}y_{a+3} ) + y_{r'-3}y_{r'-2}y_{r'-1}y_{r'} + 2y_{r'-2}y_{r'-1}y_{r'}^2 \nonumber \\
& & -S(-2\pi i \tau + Y_0+Y_1+2\sum_{a=2}^{r'} Y_a)\,.\eear
After solving the equation for $Y_a$, and replacing the value of
$Y_a$ in the potential, we obtain the glueball superpotential in a
series to order $S^4$
\be \CW_{A_{2r'-1}^{(2)} }(S;\tau) = \CW^{VY}_{A_{2r'-1} } +
h(3S^2-14 S^3+115 S^4) + {\cal O}(S^5)\,. \ee
Again this series matches exactly with that of the theory with
untwisted affine algebra $A_{r=2r'-1}^{(1)}$ above.

\subsection{$ A_2^{(1)}$ vs. $A_2^{(2)}$}

In the present computation, we are comparing twisted and untwisted
theories to quartic order in $S$. Because of this, the case of
$A_r$ with $r\le 3$ deserves special attention: the pure instanton
correction that we found in previous section begins to enter at
order $S^h$, which means that this correction is important for our
comparison in the case of $A_2$ and $A_3$. To check that our
proposal of exact Coulombic superpotential is correct, we must be
more careful to include this effect.

The superpotential for the theory with $A_2= SU(3)$ and periodic
boundary condition is
\be \CW_{A^{(1)}_2} = \wp(\balpha_0\cdot \bX)+ \wp(\balpha_1\cdot
\bX)+ \wp(\balpha_2\cdot \bX) \label{su3case}\,.\ee
It is an even double periodic function of a two dimensional vector
$\bX$ with the proper $SL(2,Z)$ symmetry. It has five stationary
points modulo Weyl symmetry, four of which are single zeros and
describe massive vacua and one of which is a double zero and so
describes massless vacua~\cite{Dorey:1999sj}. Note that
$\balpha_1$ and $\balpha_2$ are simple roots and $\balpha_0=
-\balpha_1-\balpha_2$ is the lowest negative root . We introduce
three variables, $Y_1 = \balpha_1\cdot \bX$, $Y_2=
\balpha_2\cdot\bX$ and $Y_0=2\pi i \tau + \balpha_0\cdot \bX$. We
assume that $y_a\equiv e^{Y_a}$ are of order $q^{1/3}$, and note
that $q = y_1y_2y_3$. The series expansion of the potential to
$q^2$ order becomes
\bear \CW_{A_2^{(1)}} &=& \sum_{a=0}^2(y_a +2y_a^2 +3y_a^3+ 4y_a^4
+ 5y_a^5 +
6y_a^6) \nonumber \\
& & + \sum_{a=0}^2\{y_ay_{a+1} + 2(y_ay_{a+1})^2 +
3(y_a y_{a+1})^3\} \nonumber \\
& & + y_0y_1y_2(y_0+y_1+y_2) + y_0y_1y_2 (y_0y_1 + y_1 y_2 +y_0
y_1) \nonumber \\
& & - 6y_0y_1y_2 - 18 (y_0y_1y_2)^2 -S(-2\pi i \tau +
Y_0+Y_1+Y_2)\,. \eear
Notice that this superpotential has also contributions from pure
instantons without any magnetic charge at order $q$ and $q^2$.

We can integrate over $Y_a$ and obtain the value of $Y_a$ in the
series expansion in $S$, which can be regarded as of order
$q^{1/3}$. Inserting the values of $Y_a$ back to the
superpotential, we obtain the ${\cal W}_{VY}$ potential plus the
corrections. The resulting glueball potential for $SU(3)$ in
series to order $S^6$ is
\bear \CW_{A_2^{(1)}}(S;\tau)&=& \CW^{VY}_{A_2} +3\bigl(3S^2-17
S^3+169 S^4 - \frac{4221}{2}S^5+ \frac{150336}{5} S^6 \bigr) +
{\cal O}(S^7)\,.\nonumber\\ \label{su3series} \eear

Turning to the twisted case, $A_2^{(2)}$, our proposal for the
superpotential is
\be \CW_{A_2^{(2)}} = \wp(\balpha_0\cdot \bX) + \wp(\balpha_0\cdot
\bX+ \pi i ) +\wp (\balpha_0\cdot \bX + \pi i \tau )\,,
\label{su3t} \ee
where $\balpha_0 \cdot\bX = -\sqrt{2}X$ is defined on complex
plane. One can argue that there are five gauge inequivalent vacua
of this potential, whose characteristics should be identical to
the periodic case with the potential (\ref{su3case}). In the
series expansion, we note that $\frac{2\pi i \tau}{2} -\sqrt{2} X=
Y_0$ and $2\sqrt{2}X= Y_1$. Thus, each terms in the above
potential has the series expansion as in $A_{2r}^{(2)}$ case in
subsection (6.2). Together, the twisted superpotential can be
expanded to $q^2$ order to be
\bear \CW_{A_2^{(2)}} &=& y_0 + 2y_0^2 + 3y_0^3 + 4y_0^4+5y_0^5 +
6y_0^6+ 4(
y_1+2y_1^2+3y_1^3+4y_1^4+5y_1^5+6y_1^6) \nonumber \\
& & + y_0y_1 + 2(y_0y_1)^2 + 3(y_0y_1)^3 + y_0^3y_1+ y_0^3y_1^2 +
4y_0^4y_1 - 6y_0^2y_1 - 18 y_0^4y_1^2
\nonumber \\
& & - S(-2\pi \tau +2Y_0 + Y_1)\,. \eear
The purely instantonic contribution is $-6y_0^2y_1 -18
(y_0^2y_1)^2$ in the above expression. After integration over
$Y_a$, one obtains the exactly identical series (\ref{su3series})
of the glueball superpotential. The $SL(2,Z)$ symmetry of the
superpotential for the twisted case is crucial for this
equivalence of the glueball superpotential. While we have worked
out the series to order $S^6$, we expect that the equivalence
persists to all order. It is interesting to see whether it is the
case.

\subsection{$D_{r}^{(1)}$ vs. $D_{r}^{(2)}$}

For $D_{r=r'+1}^{(2)}$, we have $\CG_0= B_{r'}$ whose root system
is studied in Sec.~(4,4). The simple roots are
\be \bbeta_0 = -\bbe_1, \;\; \bbeta_a = \bbe_a-\bbe_{a+1}, \;
a=1,2,...(r'-1), \;\; \bbeta_{r'}= \bbe_{r'}\,. \ee
With the parameters $Y_0= 2\pi i \tau +\bbeta_0 \cdot \bX$ and
$Y_a= \bbeta_a^*\cdot \bX, a=1,2,...r'$ and $y_a=e^{Y_a}$, the
superpotential (\ref{d2r}) in series expansion becomes
\bear \CW_{D_{r'+1}^{(2)}} &=& 2(y_0+2y_0+3y_0^3+4y_0^4) \nonumber \\
& & + \sum_{a=1}^{r'-1} (y_a+ 2y_a^2+3y_a^3 + 4 y_a^4) +
2(y_{r'}+2y_{r'}^2 + 3y_{r'}^3 +
4y_{r'}^4) \nonumber \\
& & + \sum_{a=0}^{r'-1}( y_ay_{a+1} + 2y_a^2y_{a+1}^2) + 2y_0y_1^2
\nonumber \\
& & + \sum_{a=0}^{r'-2} y_ay_{a+1}y_{a+2} +2 y_{r'-1}^2 y_{r'} +
y_0y_1^2y_2 \nonumber \\ & & + \sum_{a=0}^{r'-3}
(y_ay_{a+1}y_{a+2}
y_{a+3}) + y_{r'-2}y_{r'-1}^2y_{r'} \nonumber \\
& & - S(-2\pi i \tau + Y_0 + 2 \sum_{a=1}^{r'-1} Y_a + Y_{r'})
\,.\eear
Integration over $Y_a$ leads to the glueball superpotential in a
series expansion. The series to order $S^4$ becomes
\bear \CW_{D_{r'+1}^{(2)}}(S;\tau) &=& \CW^{VY}_{D_{r'+1}} +
12(r'-1)S^2- 56(2r'-3)S^3+ 368(5r'-9) S^4
\nonumber \\
&=& \CW^{VY}_{D_{r'+1}} + 3h_2 S^2 - 14 h_3 S^3 + 115 ( h_4 -
\frac{4}{5}) S^4 + {\cal O}(S^5) \label{gluedr}\,, \eear
where $h_n = \sum_{a=0}^{r'+1} (k_a^*)^n$ with the comarks $k_a^*$
of the untwisted Lie algebra $D_{r'+1}^{(1)}$. This series matches
the glueball superpotential (\ref{DE3glue}) for the untwisted
$D_{r=r'+1}^{(1)}$ case.

As in the case of $A_r$ series, the above comparison is correct
for large $h$ so that the pure instanton effect does not show up
to order $S^4$. The theory with $D_3^{(2)}$ should be compared
with the untwisted theory of gauge group $D_3(=A_3)=SU(4)$ with
$h=4$. To order $S^4$, the glueball superpotential becomes
\be \CW_{D_3^{(2)} } (S;\tau)= \CW^{VY}_{D_3 } + 4\bigl( 3 S^2 -
14 S^3 + (115 - \frac{5}{2}) S^4\bigr) + {\cal O}(S^5) \,,\ee
which matches that of the theory with untwisted $A_3=SU(4)$ gauge
group. The coefficient in $S^4$ differs from the ordinary
expansion by a pure instanton correction. This again shows that
the $SL(2,Z)$ fixes correctly the pure instanton correction.

\subsection{$E_6^{(2)}$ and $D_4^{(3)}$}

Let us first consider the $E_6^{(2)}$ case. The $\CG_0$ of the
even generators of $E_6$ are $F_4$. The simple roots of
$E_6^{(2)}$ are
\be \bbeta_0 = -\bbe_1, \;\; \bbeta_1=
\frac{\bbe_1-\bbe_2-\bbe_3-\bbe_4}{2} , \;\; \bbeta_2=\bbe_4,\;\;
\bbeta_3= \bbe_3-\bbe_4,\;\; \bbeta_4= \bbe_2-\bbe_3\,. \ee
Its dual Coxeter number is $h=12$. With $Y_0=2\pi i
\tau+\bbeta_0^*\cdot \bX$ and $Y_a=\bbeta_a^*\cdot \bX$, we see
that the superpotential (\ref{e62}) in series is
\bear \CW_{E_6^{(2)}} &=& 2\sum_{a=0}^2 (y_a
+2y_a^2+3y_a^3+4y_a^4) + \sum_{a=3}^4 (y_a +2y-a^2+3y_a^3+4y_a^4)
\nonumber \\ & & + 2\sum_{a=0}^1 (y_ay_{a+1}+ 2y_a^2 y_{a+1}^2) +
\sum_{a=2}^3 (y_ay_{a+1} + 2y_a^2y_{a+1}^2) \nonumber \\ & & +
2y_0y_1y_2 + y_1y_2y_3+ y_2y_3y_4 + + 2y_2y_3^2 \nonumber \\
& & +y_0y_1y_2y_3+y_1y_2y_3y_4+2y_1y_2y_3^2+y_2y_2^2y_4 \nonumber
\\ & &
-S(-2\pi i \tau + Y_0+2Y_1+3Y_2+4Y_3+2Y_4) \,.\eear
The glueball superpotential is
\begin{eqnarray}
\CW_{E_6^{(2)}}(S;\tau)&=& \CW^{VY}_{E_6} + 72 S^2-756 S^3+ 1490
S^4  \\
&=& \CW^{VY}_{E_6} + 3h_2S^2-14h_3S^3 + (115 h_4 - \frac{23}{2}
\hat{C}_{ab}(k_a^*)^2(k_b^*)^2 )S^4+O(S^5)\,,\nonumber
\end{eqnarray}
since $h_2= 24$, $h_3=54$, $h_4=132$, and
$\hat{C}_{ab}(k_a^*)^2(k_b^*)^2= 24$. This matches exactly with
the glueball superpotential (\ref{DE3glue}) for $E_6^{(1)}$ above.

For $D_4^{(3)}$ case, the subalgebra $\CG_0$ of $D_4^{(3)}$ is
$G_2$. The simple roots of $D_4^{(3)}$ are
\be \bbeta_0 = -\sqrt{\frac{2}{3}}\bbe_2\,, \;\;\bbeta_1 =
-\sqrt{\frac{1}{2}}\bbe_1+ \sqrt{\frac{1}{6}}\bbe_2\,,
\;\;\bbeta_2 = \sqrt{2}\bbe_1 \,,\ee
The dual Coxeter number is $h=6$, which is identical to that of
$D_4=SO(8)$. With the new variables $Y_0=2\pi i \tau
+\bbeta_0\cdot \bX$, $Y_1 = \bbeta_1\cdot\bX$ and $Y_2 =
\bbeta_2\cdot\bX$, and $y_a=e^{Y_a}$, we get the series expansion
of the superpotential (\ref{d34}) to order $q^{4/h}$;
\bear \CW_{D_4^{(3)}} &=& 3(y_0+2y_0^2+3y_0^3+4y_0^4) +
3(y_1+2y_1^2+3y_1^3+4y_1^4) + y_2+2y_2^2+3y_3^3+4y_4^4 \nonumber
\\ & &
+3(y_0y_1+ 2y_0^2y_1^2) +(y_1y_2+2y_1^2y_2^2) + y_1y_2^2 +
3y_1y_2^3 + y_0y_1y_2 +y_0y_1y_2^2 \nonumber \\ & & - S(-2\pi i +
Y_0+2Y_1+ 3Y_2) \,.\eear
Integrating over the variables $Y_a$ leads to the glueball
superpotential to order $S^4$,
\be \CW_{D_{4}^{(3)}}(S;\tau)= \CW^{VY}_{D_4} + 3h_2 S^2-141h_3
S^3+ 115(h_4 - \frac{4}{5}) S^4 + {\cal O}(S^5)\,, \ee
where $h_2= \sum_a (k_a^*)^2= 8$, $h_3= \sum_a (k_a^*)^3 = 12$,
and $h_4= \sum_a(k_a^*)^4 = 20$. This matches exactly with the
superpotential (\ref{DE3glue}) for $D_4^{(1)}$.

\section{Conclusion}

In this work we computed the superpotential of the four
dimensional $\CN=1^*$ theory compactified on $S^1$ with twisted
boundary condition. The twisting is done with help of outer
automorphisms, and because of this, twisted theories arise from
twisting of simple-laced groups only. For the case of simple-laced
classical Lie algebra, we employed an M theory construction to
argue that all the $\CN=4$ supersymmetric counterparts have the
$SL(2,Z)$ symmetries.  This $SL(2,Z)$  symmetry is expected  to be
inherited by the corresponding $\CN=1^*$ theories and acts on the
superpotential as modular transformations. We assume that the same
holds for $E$ type as well, and used this symmetry to determine
the exact superpotential for the twisted case. We explored both
symmetry aspects and vacuum physics of these superpotentials, and
find no discrepancy.

As the superpotentials do not depend on the compactification
radius explicitly, vacuum physics of the gaugino condensate should
be unaffected by the twisting. This is partially confirmed by
studying the dual glueball superpotentials in the series expansion
in the weak coupling regime, whereby we find identical results
regardless of twisting or no twisting, adding weights to our
proposal of exact superpotential.

As in the case of untwisted theories previously studied, the
superpotential for the $\CN=1^*$ theory with twisted boundary
condition is associated with the elliptic Calogero-Moser model. As
the physics described by two potentials are identical in the
$\CN=1^*$ theories, such a unity may show up in other context too.
Their spectral curves may be identical once the right
representation is chosen as the spectral curves are supposed to
describe the $\CN=2^*$ physics. There may be a more direct way to
obtain the superpotential for the twisted case from the untwisted
case by some sort of reduction. The vacua of the  $\CN=2$ theories
compactified on a circle are characterized by  hyper-K\"ahler
spaces~\cite{Seiberg:1996nz}, which may have a similar reduction.

While we have studied here the $\CN=1^*$ theories with twisted
boundary condition, one could imagine general $\CN=1$ theories,
say with different representations or different interaction, or
more general potential, with twisted boundary condition. The full
superpotential of the glueball superfield must still be
insensitive to the boundary conditions, we anticipate. It would be
interesting to check this explicitly. A method to find the
superpotential of the compactified theories on a circle has been
proposed in Ref.~\cite{Boels:2003fh}, which may be useful along
this direction.

Finally, there have been some studies of five dimensional $\CN=1$
Yang-Mills  theories compactified on a
circle~\cite{Nekrasov:1996cz,Braden:1999zp}. There vacuum
structure is closely associated with one-parameter generalization
of the Calogero-Moser models~\cite{Ruijsenaars:vq}. It would be
interesting to generalize our consideration to this case also.

\vskip 1cm

\centerline{\large\bf Acknowledgements} \vskip 5mm \noindent
H.-U.Y. and S.K. are grateful to Physics Department, University of
Texas for hospitality, and S.K. also thanks Chaiho Rim for useful
conversations. S.K. is supported in part by BK21 project of the
Ministry of Education, Korea, and also by a 2003 Interdisciplinary
Research Grant of Seoul National University. K.L. is supported by
NSF under Grant No. 0071512, K.L. and H.-U.Y are supported in part
by grant No. R01-2003-000-10391-0 from the Basic Research Program
of the Korea Science \& Engineering Foundation. P.Y. is supported
in part by Korea Research Foundation Grant KRF-2002-070-C00022.

\vskip 1cm

\startappendix

\Appendix{Glueball Superpotential of $B_r, C_r, G_2, F_4$
Theories}

While our task, that is, the comparison of glueball
superpotentials for twisted and untwisted cases, is now complete,
let us also record those of other untwisted case for the sake of
completeness. \vskip 1cm \leftline{\bf Case of $B_r$} For the
$B_r= SO(2r+1)$ gauge theory, the simple roots of the affine
algebra are
\be \balpha_0 = -(\bbe_1+\bbe_2),\; \balpha_a= \bbe_a-\bbe_{a+1} ,
a=1,2,...r-1, \;\; \balpha_r = \bbe_r \,,\ee
with $h=2r-1$. The superpotential (\ref{untpot}) in weak coupling
expansion is
\bear \CW_{B_r^{(1)} } &=& \sum_{a=0}^{r-1}(y_a
+2y_a^2+3y_a^3+4y_a^4) + 2(y_r +2y_r^2+3y_r^3+4y_r^4) +y_0y_2 +
2y_0^2y_2^2
\nonumber \\
& & + \sum_{a=1}^{r-1} (y_ay_{a+1} + 2y_a^2y_{a+1}^2)
+y_0y_1y_2+y_0y_2y_3+ \sum_{a=1}^{r-2} y_ay_{a+1}y_{a+2} +
2y_{r-1}^2y_r
\nonumber \\
& & + y_0y_1y_2y_3+y_0y_2y_3y_4 +
\sum_{a=1}^{r-3}y_ay_{a+1}y_{a+2}y_{a+3} + y_{r-2}y_{r-1}^2y_r
\nonumber \\
& & -S(-2\pi i\tau+ Y_0+ Y_1 + 2\sum_{a=1}^{r-1} Y_a + Y_r
)\,.\eear
Integrating over $Y_a$'s, the glueball superpotential becomes
\be \CW_{B_r^{(1)} } (S;\tau)= \CW^{VY}_{B_r} +(12r- 18) S^2
-(112r-224)S^3+ (1804r-4232)S^4+{\cal O}(S^5) \,.\ee

\vskip 1cm \leftline{\bf Case of $C_r$} For the $C_r=Sp(2r)$ gauge
theory, the simple roots of the affine algebra are
\be \balpha_0 = -\sqrt{2}\bbe_1, \balpha_a=
\frac{\bbe_a-\bbe_{a+1}}{\sqrt{2}}, \; a=1,2,...r-1, \;\;\;
\balpha_r = \sqrt{2}\bbe_r \,.\ee
The superpotential in weak coupling expansion to order
$q^{\frac{4}{h}}$ is
\bear \CW_{C_r^{(1)}} &=& y_0+2y_0^2+3y_0^3+4y_0^4 +
2\sum_{a=1}^{r-1} (y_a+ 2y_a^2+3y_a^3+4y_a^4) \nonumber \\
& & + y_r+2y_r^2+3y_r^3+4y_4^4 + y_0 y_1 +2y_0^2y_1^2
+ 2\sum_{a=1}^{r-2}(y_ay_{a+1}+2y_a^2y_{a+1}^2) \nonumber \\
& &+ y_{r-1}y_r +
2y_{r-1}^2y_r +y_0y_1y_2 + 2\sum_{a=1}^{r-3} y_a y_{a+1} y_{a+2} \nonumber \\
& & + y_{r-2}y_{r-1}y_r + 2y_0^2y_1 + 2y_{r-1}y_r^2 +y_0y_1y_2y_3
+ 2\sum_{a=1}^{r-4}y_a y_{a+1} y_{a+2} y_{a+3}
\nonumber \\
& & + y_{r-3}y_{r-2}y_{r-1}y_r + 2y_0^2y_1y_2 +
2y_{r-2}y_{r-1}y_r^2 -S(-2\pi i \tau+ \sum_{a=0}^r Y_a )\,, \eear
with the dual Coxeter number $h=r+1$. After integrating over the
$Y_a$, we obtain the glueball superpotential
\be \CW_{C_r^{(1)}}(S;\tau) = \CW^{VY} + (\frac{3r}{2}+3)S^2
-(\frac{7r}{2} + \frac{35}{4}) S^3 + (\frac{115}{8}{r} +
\frac{161}{4})S^4 +{\cal O}(S^5)\,.\ee

\vskip 1cm \leftline{\bf Case of $G_2$} The simple roots of
$G_2^{(1)}$ are
\be \balpha_0 = -\sqrt{\frac{1}{2}}\bbe_1
-\sqrt{\frac{3}{2}}\bbe_2 , \; \balpha_1 = \sqrt{2}\bbe_1, \;
\balpha_2= -\sqrt{\frac{1}{2}}\bbe_1 + \sqrt{\frac{1}{6}}\bbe_2\,.
\ee
The Coxeter number is $h=4$ as $\balpha_0 +2\balpha_1+ \balpha_2^*
= 0 $. The superpotential in weak coupling expansion is
\bear \CW_{G_2^{(1)}} &=& \sum_{a=0}^1 (y_a +2y_a^2+3y_a^3+4y_a^4)
+ 3(y_2 +
2y_2^2+ 3y_2^3 +4y_4^4) + y_0y_1 + 2y_0^2y_1^2 \nonumber \\
& & + y_1y_2 + 2y_1^2y_2^2 + +y_0y_1y_2+y_1^2y_2+3y_1^3y_2-6y_0y_1^2y_2 \nonumber \\
& & -S(-2\pi i \tau+ Y_0+2Y_1+Y_2)\,.\eear
Here the term $-6y_0y_1^2y_2 $ is due to the contribution of a
pure single instanton. The glueball superpotential is
\be \CW_{G_2^{(1)}}(S;\tau) = \CW^{VY}_{G_2}
+\frac{40}{3}S^2-\frac{700}{9}S^3+\frac{66632}{81}S^4+O(S^5)
\,.\ee

\vskip 1cm \leftline{\bf Case of $F_4$} The simple roots of
$F_4^{(1)}$ are
\be \balpha_0=-\bbe_1-\bbe_2, \;\balpha_1 =\bbe_2-\bbe_3 ,\;
\balpha_2= \bbe_3-\bbe_4, \; \balpha_3= \bbe_4,\; \balpha_4 =
\frac{\bbe_1-\bbe_2-\bbe_3-\bbe_4}{2} \,.\ee
The dual Coxeter number is $h= 9$. With $Y_0 = 2\pi i \tau+
\balpha_0\cdot \bX$ and $Y_a = \balpha_a^*\cdot \bX$, the
superpotential becomes
\bear \CW_{F_4^{(1)}} &=& \sum_{a=0}^2 (y_a+ 2y_a^2+3y_a^3+4y_a^4)
+ 2 \sum_{a=3}^4 (y_a+ 2y_a^2+3y_a^3+4y_a^4) + \sum_{a=0}^2
(y_ay_{a+1}+ 2y_a^2y_{a+1}^2)
\nonumber \\
& & + 2(y_3y_4+ 2y_3^2y_4^2) + y_0y_1y_2+ y_1y_2y_3+ 2y_2^2y_3 +
y_2y_3y_4+ y_0y_1y_2y_3
\nonumber \\
& &+ y_1y_2^2y_3+ y_1y_2y_3y_4 + 2y_2^3y_3y_4 -S(-2\pi i \tau +
Y_0+2Y_1+3Y_2+2Y_3+Y_4)\,. \eear
The glueball superpotential for $F_4^{(1)}$ becomes
\be \CW_{F_4^{(1)}} (S;\tau)= \CW^{VY}_{F_4}
+45S^2-\frac{1575}{4}S^3+\frac{25875}{4}S^4+{\cal O}(S^5)\,. \ee

\newpage
\Appendix{Lie Algebra}

We choose an orthonormal basis $\{ T^i\} $ of a simple Lie algebra
${\cal G}$ of rank $r$ with respect to the group invariant inner
product defined in the adjoint representation,
\be \Tr(T^i, T^j) = \frac{1}{C_2(\CG)} \tr\; T^i T^j =
\delta^{ij}\,, \ee
with a normalization coefficient $C_2(\CG)$. (Here $\tr$ is a
matrix trace and $\Tr$ is a just notation for the inner product
defined as above.) On this basis, the second Casimir operator is
$C_2(\CG)=\sum_i (T^i)^2$ in the adjoint representation. A maximal
set $\bH=(H_1,H_2,...H_r)$ of mutually commuting generators $H_a$,
where $a=1,2,...r$, forms an $r$-dimensional vector. The rest of
generators can be combined as step operators $E_\balpha$ such that
\be [\bH, E_\balpha] = \balpha E_\alpha \label{becomm}\,, \ee
with the root vector $\balpha$ in $r$-dimension. With its
hermitian dual $E_{-\alpha}= E_\alpha^\dagger$, it satisfies
\be [E_\balpha, E_\balpha^\dagger] = \balpha^*\cdot \bH
\label{ecomm}\,, \ee
where $\balpha^* $ is the co-root defined by
\be \balpha^* = \frac{2\balpha}{\balpha^2} \,.\ee
We choose the normalization so that the longest roots have the
length square two. It is known that the normalization coefficient
becomes
\be C_2(\CG) = 2h\,, \ee
with the dual Coxeter number $h$ defined later.

There is a set of simple roots $\balpha_a\,, \; a=1,2,...r$ which
spans the root lattice $\Lambda_R$. The lowest negative root is
denoted by $\balpha_0$, which turns out to be a long root. With
our normalization $\balpha_0^2=2$. The co-roots of the simple
roots span the co-root lattice $\Lambda_R^*$. The weight lattice
$\Lambda_W$ is spanned by fundamental weights $\bw_a$ satisfying
$\bw_a\cdot \balpha_b^*=\delta_{ab}$. The co-weight lattice
$\Lambda_W^*$ is spanned by the fundamental co-weights $\bw_a^*=
(2/\balpha_a^2) \bw_a$. For any simple Lie algebra, the length of
roots have at most two different values. Our normalization implies
that any root in a simply laced algebra has length square two, and
any long root in a non-simply laced algebra has length square two.
The extended Dynkin diagram of Lie algebra $\CG$ is defined by
$\balpha_a, a=0,1,...r$ and identical to the Dynkin diagram of the
corresponding untwisted affine algebra $\CG^{(1)}$ which is
discussed in the next appendix. They are shown in Figure 3, where
the solid red dots denote $\bbeta_0$.

There exists a unique set of positive integers $k_a^*$, the
so-called co-marks, such that $k_0^*=1$ and
\be \sum_{a=0}^r k_a^* \balpha_a^* = 0\,. \ee
It is minimal in the sense that there is no smaller positive
integers such that the above equation is true. The dual Coxeter
number is defined as
\be h =\sum_{a=0}^r k_a^*\,,\ee
which is always bigger than the rank $r$ of the algebra. These
comarks for the corresponding roots are shown in the Figure 3.

For a given root $\balpha$, a $SU(2)$ subalgebra is defined by
three generators,
\be t^1=\frac{1}{2}(E_\alpha+E_\alpha^\dagger), \;\; t^2 =
\frac{1}{2i}(E_\alpha-E_\alpha^\dagger),\;\; t^3 =
\frac{1}{2}\balpha^*\cdot \bH \,,\ee
which satisfy the $SU(2)$ algebra $[t^i,t^j]=i\epsilon^{ijk}t^k$.

The Weyl vector $\rho =\sum_{i=1}^r \bw_i$ and the level vector
$\rho^*=\sum_{i=1}^r \bw_i^*$ of a Lie algebra ${\cal G}$ are
related to the set of all positive roots, ${\cal R}_+({\cal G})$,
by the following equations,
\bear && \brho = \sum_{i=1}^r \bw_i = \frac{1}{2}\sum_{\balpha \in
{\cal R}_+({\cal G})} \balpha\,, \nonumber \\
&& \brho^* =\sum_{i=1}^r \bw_i^* = \frac{1}{2} \sum_{\balpha\in
{\cal R}_+({\cal G}) } \balpha^*\,. \eear
The level $\l(\lambda)$ and co-level $\l^*(\lambda)$ of a weight
$\lambda$ are defined by
\be \l(\lambda) = \lambda\cdot \brho^*, \;\;\; \l^*(\lambda)=
\lambda\cdot \brho \,,\ee
respectively. All roots take the integer levels and all co-roots
the integer co-levels. Note that the dual Coxeter number is
related the co-level of $\balpha_0^*$,
\be h = 1- \balpha_0^*\cdot\rho \,.\ee
While $\balpha_0$ has the lowest level, its coroot $\balpha_0^*$
does not have the lowest co-level, or the lowest root of the
co-root system. However the values
\be \balpha^*\cdot \rho -\frac{2h}{\balpha^2}\,, \ee
for all roots $\balpha$ takes the maximum value one for the
positive root $-\balpha_0$.

\Appendix{Untwisted Affine Lie Algebra}

We start with a simple Lie algebra $\CG$ of rank $r$ with
generators $T^i$. By attaching integer-valued gradings $n$ to
these generators, we find the corresponding untwisted affine Lie
algebra $\CG^{(1)}$ of infinite number of generators $T^i_n$,
which satisfies
\be [T^i_m,T^j_n]\,=\,i
f^{ijk}T^k_{m+n}\,+\,pn\delta^{ij}\delta_{m+n,0}\,. \ee
The constant $p$ is a central charge which commutes with all
elements of $\CG^{(1)}$, and it will vanish in all cases of our
present interest. From the Cartan subalgebra of $\CG$, we choose
the corresponding commuting operators $(H_0^1,H_0^2,...H_0^r)$ in
the zero-grading. In addition we introduce the operator $d$ which
measures the grading,
\be [d, T^i_n] = nT^i_n\,. \ee
Then roots of the untwisted affine Lie algebra $\CG^{(1)}$ are
$(r+1)$-component vectors composed of the eigenvalues of
$\{\,H^a_0,\,d\}$. It is not difficult to check that a simple root
system of $\HCG$, of which every root is either all positive or
all negative linear combination, is provided by
\be (\balpha_a,0),\,a=1,\ldots r, \,\,\,\, \,(\balpha_0,1) \,,\ee
where $\balpha_a$ ($\balpha_0$) are simple roots (the lowest
negative root) of $\CG$. The Dynkin diagram shown in Figure 3 of
the untwisted affine algebra $\CG^{(1)}$ is made of
$\{\balpha_a\}$, $i=0,1,\ldots,r$ and $\balpha_0$, and so
identical to the extended Dynkin diagram of the original Lie
algebra $\CG$. For each positive root $\balpha$ and grading $n\in
Z$, there exist a raising operator $E_\alpha^n$ and a lowering
operator $(E_\alpha^n)^\dagger =E_{-\alpha}^{-n}$. With
$\balpha^*\cdot \bH_0$, these operators define three generators
for the $SU(2)$ algebra,
\be t^1=\frac{1}{2}(E^n_\alpha+(E^n_\alpha)^\dagger), \;\; t^2 =
\frac{1}{2i}(E^n_\alpha-(E^n_\alpha)^\dagger),\;\; t^3 =
\frac{1}{2}\balpha^*\cdot \bH_0 \,.\ee

\vspace{1em}

\Appendix{Twisted Affine Lie Algebra}

A twisted affine Lie algebra, $\CG^{(L)}$, out of an ordinary
simple Lie algebra $\CG$ of rank $r$, is based on an outer
automorphism $\bsigma$ of $\CG$ such that $\bsigma^L=1$ with the
smallest positive integer $L$. First we divide elements of the
complex vector space $\CG$ into subspaces $\CG_n$, $n=0,\ldots
L-1$, by their eigenvalues $e^{\frac{2\pi i n}{L}}$ with respect
to $\tau$. Because $\bsigma$ is an automorphism, it preserves the
group and algebraic relations. Thus,
\be \bsigma([\CG_n,\CG_{n'} ])=[\bsigma(\CG_n),\bsigma(\CG_{n'})]
=e^{\frac{2\pi i (n+n')}{L}}[\CG_n,\CG_{n'}]\,, \ee
which means that $[\CG_n,\CG_{n'}]\subset \CG_{n+n'\; {\rm mod} \;
L }$. Note that the subspace $\CG_0$ is special as it is a Lie
algebra by itself of rank $r'<r$. All other subspaces $\CG_n, n\ne
0$ form representations of the Lie algebra $\CG_0$.

The original elements of the Lie algebra can be expressed as
linear combinations of the elements $T^i_\frac{n}{L}$ of these
subspaces $\CG_n$, whose structure constant would be $f^{ijk}_{n,
n'}$ We now assign integer gradings to each of these elements of
$\CG_n$ to obtain generators $T^i_s $ with $s= \frac{n}{L} +m$,
and require them to satisfy the commutation relation of the
twisted affine Lie algebra $\CG^{(L)}$,
\be [T^i_s,T^j_{s'}]\,=\,i f_{n,n'}^{ijk}T^k_{s+s'}\,+\,p s
\delta^{ab} \delta_{s+s',0}\,. \ee
which is consistent with the original Lie algebra. The central
term, $p$, vanishes for the case we study here.

The outer automorphism $\bsigma$ modulo inner automorphism is
known to be equivalent to the exchanging operator of simple roots
$\balpha_a, a=1,2,...r$ of the Lie algebra $\CG$ according to the
symmetry of its Dynkin (not extend Dynkin) diagram. Thus the
exchange transformation $\bsigma$ defines the outer automorphism
of the raising operators of simple roots as
\be \bsigma (E_{\balpha_a}) = E_{\bsigma(\balpha_a)} \,, \ee
It is straightforward to extend this outer automorphism to the
raising operators of other positive roots as they are given by the
commutations of $E_{\balpha_a}$. Also one obtains the $\bsigma$
transformation of the lowering operators by noting
$\bsigma(E_{\balpha_a}^\dagger)=E_{\bsigma(\balpha_a)}^\dagger$.
The extension to the Cartan subalgebra can be obtained by the
commutation relation between raising and lowering operators. As we
know the transformation of all generators of $\CG$, we can
construct the eigen operator space $\CG_n$ of the outer
automorphism by linear combination of operators.

The exchange symmetry of simple roots in the Dynkin diagram can be
easily extended to the exchange symmetries of all roots. It
preserves the length of the root and the angle between the basis
$\balpha_a$. Thus it can be regarded as an element of the
orthogonal group $O(r)$. We use the same notation $\bsigma$ for
both the outer automorphism and this linear map without confusion.
{}From the outer automorphism of the commutation (\ref{ecomm}), we
get
\be ([E_{\bsigma(\balpha_a)},E_{\bsigma(\balpha_a)}^\dagger] =
\bsigma(\balpha_a^*\cdot \bH) =\bsigma(\balpha_a^*)\cdot\bH \,,\ee
which implies $\bsigma(\balpha\cdot\bH)=\bsigma(\balpha)\cdot\bH$.
This is consistent with the outer automorphism of a variation
$[\bbeta\cdot \bH, E_{\alpha_i}]= \bbeta\cdot\balpha E_{\alpha_i}$
of the commutation relation (\ref{becomm}), which is
\be [\bsigma(\bbeta\cdot\bH),E_{\bsigma(\balpha_a)}] = \bbeta\cdot
\balpha_a E_{\bsigma(\alpha_a)}\,. \ee
The left hand side of the above equation is
\be [\bsigma(\bbeta)\cdot\bH, E_{\bsigma(\alpha_i)}] =
\bsigma(\bbeta)\cdot\bsigma(\balpha_i) E_{\bsigma(\alpha_i)}\,,
\ee
which is identical to the previous equation as
$\bsigma(\bbeta\cdot\balpha)=\bbeta\cdot\balpha$.

Thus the elements of the Cartan subalgebra which belong to $\CG_0$
would be made of $\bsigma(\bzeta\cdot\bH)=\bzeta\cdot\bH$ or
$\bsigma(\bzeta)=\bzeta$. So $\CG_0$ will be a $r'$ dimensional
Lie algebra whose Cartan algebra is spanned by $\sum_{l=0}^L
\bsigma^{\l} (\balpha_a)\cdot \bH\,,a=1,2,...r $. Similarly,
$\sum_{l=0}^L e^{-2\pi i l n/L} \bsigma^l(\balpha_a)\cdot\bH\,,
i=1,2...r$ would belong to $\CG_n$. Since
$[\bH,E_{\bsigma(\alpha)}] = \bsigma(\balpha) \bsigma( E_{\alpha}
)$, we get $\bsigma(E_{\alpha}) = \epsilon(\bsigma,\alpha)
E_{\bsigma(\alpha)}$ such that $\epsilon(\bsigma^L,\alpha)=1$ as
$\bsigma^L=1$. Thus the linear combinations $\sum_{l=1}^L e^{-2\pi
i nl/L} \bsigma^l(E_{\alpha})$ would belong to $\CG_n$.

For any root vector $\balpha$, $\sum_{l=1}^L
\bsigma^l(\balpha_a)\cdot\balpha =
\sum_{l=1}^L\bsigma^l(\balpha_a)\cdot \frac{1}{L}\sum_{l=1}^L
\bsigma^l(\balpha)/L$. Thus we see that
\be [\sum_{l=1}^L \bsigma^l(\balpha_a)\cdot\bH, \sum_{l=1}^L
e^{-2\pi i nl/L} \bsigma^l( E_{\balpha}) ] = \left(\sum_{l=1}^L
\bsigma^l(\balpha_a)\cdot \frac{1}{L}\sum_{l=1}^L
\bsigma^l(\balpha) \right) \sum_{l=1}^L e^{-2\pi i nl/L}
\bsigma^l(E_{\alpha})\,. \ee
This means that the weight of root vectors of an element
$\sum_{l=1}^L e^{-2\pi i nl/L} E_{\bsigma^l(\alpha)} $ in $\CG_n$
is given in $\frac{1}{L}\sum_{l=1}^L \bsigma^l(\balpha)$.

Now we have the grading of the Lie algebra due to the outer
automorphism explicitly and so we can define the twisted affine
algebra $\CG^{(L)}$. The maximally commuting subalgebra of
$\CG^{(L)}$ would be a $r'$ vector $ \bH' = \{\frac{1}{L}\sum_l
\bsigma^l(\balpha_a)\cdot \bH_0 \}$ of $\CG_0$ and the grading
operator $d$ whose eigenvalue is $s=m+ n/L$. The simple roots of
$\CG^{(L)}$ would be
\be \left\{(\bbeta_a,0), (\bbeta_0,\frac{1}{L})\right\} = \left\{
\left( \frac{1}{L}\sum_{l=1}^L \bsigma^l(\balpha_i), 0\right) ,
(\bbeta_0,\frac{1}{L})\right\} \,,\ee
where $j=1,2,...r'$ and $\bbeta_0$ is the lowest negative weight
of $\CG_1$ with respect to $\CG_0$. The simple roots of sub Lie
algebra $\CG_0$ are $\bbeta_a\,, \; a=1,2,...r'$. These simple
roots $\bbeta_a\,, a=0,1,2,...r'$ of the twisted affine algebra
$\CG^{(L)}$ define the Dynkin diagram of the twisted affine Lie
algebra, which is shown in Figure 3.

\Appendix{Dynkin Diagrams and Some New Identities}

The nontrivial outer automorphism is possible only for simply
laced group with symmetric Dynkin diagram. For $A_r =SU(r+1),
D_{r+1}= SO(2(r+1)$ and $E_6$, the order of the symmetry is $L=2$.
For $D_4=SO(8)$, there is also the triple symmetry of order $L=3$.
For all simple Lie algebra $\CG$, their untwisted affine algebra
is denoted as $\CG^{(1)}$, and their twisted affine algebra would
be denoted as $\CG^{(L)}$. Their Dynkin diagram is given in Figure
3. Note that the lowest root $\balpha_0$ in all untwisted affine
algebra which appears as a filled circle is a long root. The root
$\bbeta_0$ in $A_{2r}^{(2)}$ case is a long root. For all other
twisted affine case the root $\bbeta_0$ is a short root.

Once we fix the normalization of the elements of the Lie algebra
$\CG$ as in Appendix B, which fixes the length of the root, the
procedure in the previous appendix fixes the normalization of the
elements of untwisted and twisted affine algebra. For our
normalization the long roots of the untwisted affine algebra would
have length square two. This leads to the fact that for all
twisted affine algebras, the longest root has square length two.

For a Lie group $\CG$ of rank $r$, its extended Dynkin diagram of
simple roots $\balpha_a\,, a=1,2...r$ and the lowest root
$\balpha_0$ is that of the corresponding untwisted affine algebra.
Similarly the simple roots $\bbeta_a\,, a=0,1,...r'$ defines the
Dynkin diagram of the twisted affine algebra $\CG^{(L)}$. One can
see that the dual simple roots $\balpha_a^*\,, a =0,1,... r$ of
untwisted affine algebra defines the dual Dynkin diagram of
untwisted affine algebra. Similarly the dual of simple roots
$\bbeta_a^*\,,a=0,1,...r'$ defines the dual Dynkin diagram of the
twisted affine algebra. The Dynkin diagrams of $B_r^{(1)}$ and
$A_{2r-1}^{(2)}$ are dual to each other. The same is true for
pairs, ($C_{r}^{(1)}$, $D_{r+1}^{(2)}$), ($F_4^{(1)}$,
$E_6^{(2)}$), and ($G_2^{(1)}$, $D_4^{(2)}$). The rest of Dynkin
diagrams in Figure 3 are self-dual.

For untwisted Lie algebra $\CG^{(1)}$ for a Lie group $\CG$ of
rank $r$, their simple roots are $\balpha_a\,, i=0,1,...r$, which
appear in the extended Dynkin diagram of $\CG$. As defined in
Appendix A, the comarks $k_a^*$ and the dual Coxeter number
$h(\CG^{(1)})$ are defined so that $k_0^*=1$, and
\be \sum_{a=0}^r k_a^* \balpha_a^*=0 , \;\;\; h (\CG^{(1)}) =
\sum_{a=0}^r k_a^* \,.\ee
For twisted Lie algebra $\CG^{(L)}$ for a Lie group $\CG$ of rank
$r$, their simple roots are $\bbeta_a\,, a=0,1,...,r'$ with
$r'<r$. One can define the co-marks $\tilde{k}_a^*$ and the dual
Coxeter number $h(\CG^{(L)})$ such that
\be \sum_{a=0}^{r'} \tilde{k}_a^* \bbeta_a^* = 0 , \;\;\;
h(\CG^{(L)} ) = \sum_{a=0}^{r'} \tilde{k}_a^* \,,\ee
where $\tilde{k}_0^*=2 $ is for $A_{2r}^{(2)}$ and $\tilde{k}_0^*
=1$ for the rest of twisted affine algebra.

\begin{figure}[!t]
\hskip 4cm
\begin{center}
\includegraphics[width=0.9\textwidth]{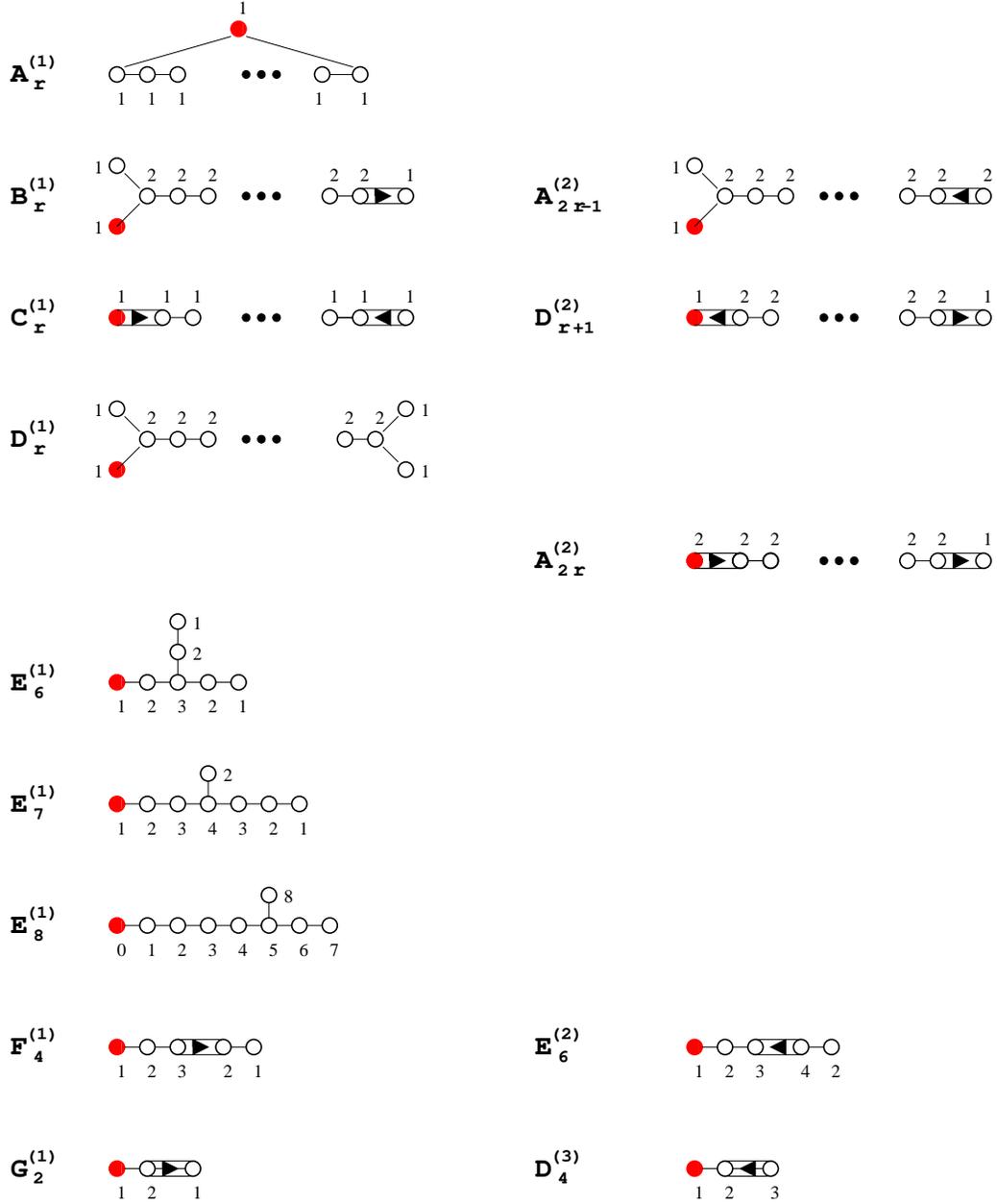}
\end{center}
\caption{Dynkin Diagrams for Affine Algebra with comarks}
\end{figure}

There are four identities for the root systems of untwisted and
twisted affine algebra we notice here. It is straight-forward to
prove them by going over the case by the case.

\begin{itemize}
\item
\noindent {\bf Fact 1 : } The dual Coxeter number does not change
with twisting. For a Lie algebra $\CG$ which has nontrivial outer
automorphism, the dual Coxeter number for the untwisted case is
identical to that of the twisted case,
\be h(\CG^{(1)}) = h(\CG^{(L)} ) \label{fact1}\,.\ee
Thus we denote the dual Coxeter number by $h$ regardless of the
twisting.

\item \noindent {\bf Fact 2 : } The above construction of simple
roots $\bbeta_a$'s from $\balpha_a$'s comes with a particular
relative normalization between generators of $\CG^{(1)}$ and those
of $\CG^{(L)}$. This choice is especially convenient because it
seems to imply a universal relationship of the form,
\be
\frac{2}{\balpha_0^2}\,\,=\,\,\frac{1}{1}\,\frac{2k^*_0}{\balpha_0^2}\,\,
=\,\,\frac{1}{L}\,\frac{2\tilde{k}_0^*}{\bbeta_0^2} \,,
\label{fact2}\ee
regardless of which Lie algebra we started with. Note that
$k^*_0=1$.

\item
\noindent {\bf Fact 3 : } With the same normalization convention,
another nontrivial and universal identity holds,
\be \prod_{a=0}^r\,\bigg[ \frac{{k}_a^*\,\balpha_a^2}{2}
\bigg]^{{k}_a^*}\,\, =\,\,\prod_{a=0}^{r'}\,\bigg[
\frac{\tilde{k}_a^* \bbeta_a^2}{2}\bigg] ^{\tilde{k}_a^*}\,.
\label{fact3} \ee
This quantity is one for the Lie algebra $A_r=SU(r+1)$.

\item
\noindent{\bf Fact 4 :} For a Lie algebra $\CG$ with outer
automorphism, the number of positive roots of the Lie algebra
$\CG$ of $\CG^{(1)}$ is related to the number of postive roots of
$\CG_0$ of the twisted affine algebra $\CG^{(2)}$ as follows. Here
we use the convention the longest roots of the (twisted) affine
algebra have length square two.
\be \#\;{\rm of}\; {\rm positive }\;{\rm roots}\; {\rm of} \; \CG
=\sum_{\rm all\; positive\; roots \bbeta \;of \;\CG_0}
\frac{2}{\bbeta^2}\;\;{\rm for}\;{\rm all}\; \CG^{(2)} \,,\ee
which holds in all cases with the exception of $A_{2r}^{(2)}$. For
the latter we have instead \be \#\;{\rm of}\; {\rm positive
}\;{\rm roots}\; {\rm of} \; \CG = 3r+ \sum_{\rm medium\; size\;
positive\; roots\; \bbeta_m \;of\; \CG_0} \frac{2}{\bbeta_m^2}\,,
\; \ee
where $\bbeta_m$'s are the medium size roots introduced in
subsection (5.2).

\end{itemize}
\vspace{1em}

\Appendix{Elliptic Functions}

The Weierstrass elliptic function is
\be \wp(z;2\pi i ,2\pi i \tau) \equiv \frac{1}{z^2} +
\sum_{m_1,m_2\in \mathbb{Z}}\!\!\!\!\!{}^{'} {}^{}\left\{
\frac{1}{(z+2\pi i m_1+ 2\pi i \tau m_2)^2} - \frac{1}{(2\pi i
m_1+2\pi i \tau m_2)^2} \right\} \,,\ee
where the prime in the sum means to exclude the point
$(m_1,m_2)=(0,0)$. This function is even under $z\rightarrow -z$
and doubly periodic under $z\rightarrow z+ 2\pi i m+2\pi i \tau n$
with $m,n \in \mathbb{Z}$. It has a double pole at $z=0$. We can
choose ${\rm Im}(\tau)>0$. We use a simpler notation
$\wp(z)=\wp(z;2\pi i , 2\pi i\tau)$ unless we have different
periods. The Weierstrass elliptic function can be represented by a
sum
\be \wp(z) = \frac{1}{12}+ \sum_{n=-\infty}^\infty \frac{1}{ q^n y
+ q^{-n} y^{-1} -2} - \sum_{n=1}^\infty \frac{2}{q^n+q^{-n}-2}\,,
\ee
where
\be q = e^{2\pi i \tau} , \;\;\; y=e^{z} \,.\ee
For small $q$ and $e^z$, we can expand the above expression in a
series
\be \wp(z) =\frac{1}{12} + \sum_{k=1}^\infty k y^k +
\sum_{k=1}^\infty \sum_{n=1}^\infty kq^{kn}(y^k+y^{-k} -2)\,. \ee
The Weierstrass elliptic function has the $SL(2,\mathbb{Z})$
symmetry under which
\be \wp(z; 2\pi i (c\tau+d) , 2\pi i (a\tau+b))=\wp(z;2\pi i ,2\pi
i \tau)\,, \ee
for all $a,b,c,d\in\mathbb{Z}$ and $ad-bc=1$. Thus the $SL(2,Z)$
transformation
\bear && \tau \rightarrow \frac{a\tau+b}{c\tau+d}\,, \\
&& z\rightarrow \frac{z}{c\tau+d}\,, \eear
leads to a modular function of weight two,
\be \wp(z;2\pi i , 2\pi i\tau) \rightarrow (c\tau+d)^2 \wp(z;2\pi
i ,2\pi i \tau)\,. \ee
The Weierstrass elliptic function satisfies an obvious homogeneity
relation,
\be \wp ( xz; 2\pi i x,2\pi i \tau x) = \frac{1}{x^2} \wp(z;2\pi i
,2\pi i \tau)\,, \ee
which allows
\be \wp(z; \pi i ,2\pi i \tau) = 4 \wp(2z;2\pi i, 4\pi i \tau)\,.
\ee
In addition it satisfies a double-period formula
\be \wp(z;\pi i,2\pi i \tau) = \wp(z) + \wp(z+\pi i) -\wp(\pi
i)\,, \ee
and a triple period formula
\be \wp(z,2\pi i /3, 2\pi i \tau) = \wp(z) + \wp(z+\frac{2\pi i
}{3}) + \wp(z+\frac{4\pi i }{3}) - \wp(\frac{2\pi i
}{3})-\wp(\frac{4\pi i }{3})\,. \ee

\newpage


 \vfil

\end{document}